# Atmospheric processes affecting methane on Mars

Grenfell, J. L.[1]
0000-0003-3646-5339
Wunderlich, F.[1,2]
0000-0002-2238-5269
Sinnhuber, M.[3]
0000-0002-3527-9051
Herbst, K.[4]
0000-0001-5622-4829
Lehmann, R.[5]
Scheucher, M.[2,6]
0000-0003-4331-2277
Gebauer, S.[1]
0000-0001-7450-3207
Arnold, G.[1]
0000-0002-2899-5009
Rauer, H.[1,2,7]
0000-0002-6510-1828

(1) Institut für Planetenforschung (PF)
Deutsches Zentrum für Luft- und Raumfahrt (DLR)
Rutherfordstr. 2
12489 Berlin
Germany

(2) Zentrum für Astronomie und Astrophysik (ZAA)
Technische Universität Berlin (TUB)
Hardenbergstr. 26
10623 Berlin
Germany

(3) Institute for Meteorology and Climate Research
Karlsruhe Institute of Technologie (KIT)
Hermann-von Helmholtz Platz 1
76344 Eggenstein-Leopoldshafen
Germany

(4) Institute of Experimental and Applied Physics
Extraterrestrial Physics
Kiel University (CAU)
Leibnitzstr. 11
24118 Kiel
Germany

(5) Alfred Wegener Institute (AWI)
Helmholtz Centre for Polar and Marine Research
Telegrafenberg A45
14473 Potsdam
Germany

(6) Now at: NASA Jet Propulsion Laboratory (JPL)
4800 Oak Grove Drive
Pasadena
CA 91109 USA

(7) Institut für Geologische Wissenschaften
Freie Universität Berlin (FUB)
Malteserstr. 74-100
12249 Berlin
Germany

*Abstract: It is currently uncertain as to whether methane exists on Mars. Data from the Curiosity Rover suggests a background methane concentration of a few tenths parts per billion whereas data from the Trace Gas Orbiter suggest an upper limit of twenty parts per trillion. If methane exists on Mars then we do not understand fully the physical and chemical processes affecting its lifetime. Atmospheric models suggest an over-estimate in the lifetime by a factor of around six hundred compared with earlier observations. In the present work we assume the Curiosity Rover background methane value and estimate the uncertainty in atmospheric chemistry and mixing processes in our atmospheric column model 1D TERRA. Results suggest that these processes can only explain a factor of ~sixteen lowering in the methane lifetime. This implies that if methane is present then additional, currently unknown processes are required to explain the observed lifetime.*

**Key words**: Mars, atmosphere, methane, lifetime, photochemical, cosmic rays

**1. Introduction**

The detection of methane ($CH_4$) in the Martian atmosphere (Krasnopolsky et al., 2004) has a fascinating history. The initially claimed abundance of several tens of parts per billion (ppbv) has steadily decreased as instrument sensitivity, analysis techniques and data coverage have improved from the initial ground-based observations on Earth (see e.g. Formisano et al., 2004; Mumma et al., 2009; Fonti and Marzo, 2010; Geminale et al., 2011; Fonti et al., 2015; Webster et al., 2018; Aoki et al., 2018; Korablev et al., 2019; Knutsen et al., 2021; Montmessin et al., 2021)). Recent overview papers on Martian methane include Krasnopolsky and Lefèvre (2013), Yung et al. (2018) and Lefèvre (2019).

There is still however an ongoing debate as to whether methane exists on Mars. Important papers which contributed to this skepticism include Lefèvre and Forget (2009) and Zahnle et al. (2011) as well as the recent Trace Gas Orbiter (TGO) observations (Korablev et al., 2019; Montmessin et al., 2021; see also discussion below). Differences in the $CH_4$ abundances measured by the Curiosity Rover (see e.g. Webster et al., 2015; Webster et al., 2018, Webster et al., 2021) and instruments onboard the TGO (Korablev et al. (2019); Knutsen et al. (2021); Montmessin et al. (2021)) are the subject of discussion. The large $CH_4$ emissions suggested by some works would imply a large sink which is not supported by current photochemical schemes.

A central aspect is the discussion on reported seasonal variation. The Mars Science Laboratory (Webster et al., 2018; see also Yung et al., 2018) suggested a baseline level of 0.4 part per billion by volume (ppbv) $CH_4$ with seasonal behavior and spikes of up to 7 ppbv. The Curiosity Rover at the Gale Crater (4.5°S,137.4°E) (Webster et al., 2018) suggested an annual cycle ranging around from 0.24 to 0.65 ppbv $CH_4$ (mean 0.41 ppbv) peaking in Northern Summer, superimposesd with episodic spikes of around 7 ppbv typically lasting a few weeks. Moores et al. (2019) and Viúdez-Moreiras et al. (2020) suggested that the annual cycle variations could be explained by regolith diffusion or/and micro seepage. However, Gillen et al. (2020) applied Gaussian regression to the same Curiosity Rover dataset but their result did not support an annual variation. Data from the Planetary Fourier Spectrometer (PFS) (Giuranna et al., 2019) supported a $CH_4$ plume of 15.5 ppbv above the Gale crater. Korablev et al. (2019), however, did not detect any $CH_4$ over a range of latitudes using the Trace Gas Orbiter (TGO) which has the highest sensitivity (down to ~0.01 ppb $CH_4$) applied to date. Montmessin et al. (2021) suggested a $CH_4$ upper limit of 20 pptv.

Initial observations from the TGO with the Atmospheric Chemistry Suite (ACS) and with the Nadir and Occultation for MArs Discovery (NOMAD) suggested an absence of $CH_4$ detection over most of the Martian globe. Abundant dust loadings in the lower atmosphere (<30 km) however sometimes made such observations challenging. Knutsen et al. (2021) also reported no confirmed $CH_4$ detection with a sensitivity down to 0.06 ppbv based on near global observations over one year from 6-100 km with the ACS and NOMAD (Liuzzi et al., 2019) instruments. Zahnle and Catling (2019) noted that claimed $CH_4$ abundances since 2004 have decreased with instrument sensitivity, suggesting that the claimed detections could be erroneous. Olsen et al. (2020) presented the first mid-IR detection of Martian ozone ($O_3$). Better $O_3$ constraints suggest improved understanding of the Martian photochemical environment which could lead to improved knowledge of $CH_4$ photochemistry. Their study noted that overlapping of the v3 vibration-rotation $CH_4$ band with a newly-discovered magnetic dipole band of $CO_2$ in the region from (3000-3060) $cm^{-1}$ could have led to an overestimation of $CH_4$ by the Curiosity Rover and PFS studies. This claim, however, was refuted by Webster et al. (2020) for the Curiosity data and more quantification is needed.

The $CH_4$ sources and sinks on Mars are not well constrained although this is a pre-requisite for assessing whether the claimed $CH_4$ signals could have arisen from life. Suggested sources of Martian $CH_4$ are geology (Oze and Sharma, 2005; Lyons et al., 2005; Chastain and Chevrier, 2007; Thomas et al., 2009; Komatsu et al., 2011; Oehler and Etiope, 2017; Etiope et al., 2018) as well as clathrates (Chassefièrre, 2009; Thomas et al., 2009; Mousis et al., 2016); delivery, including comets (Kress and McKay, 2004; Fries et al., 2016 (but contested by Roose-Serota et al., 2016)) and (micro)meteorites (Krasnopolsky et al., 2004; Court and Sephton, 2009; Keppler et al., 2012; Blamey et al., 2015; Civiš et al., 2019) and the effect of dust (Schuerger et al., 2012; Moores et al., 2017) and biological microorganisms (e.g. Summers et al., 2002; Buford Price, 2010; Nixon et al., 2013; Westall et al., 2015; Sholes et al., 2019). Wong et al. (2004) suggested that $CH_4$ could form ethane ($C_2H_6$) via gas-phase reactions in the atmosphere; in theory, the $C_2H_6$ could then decompose in-situ to reform $CH_4$ although this process is uncertain in the Martian atmosphere. Pla-Garcia et al. (2019) applied a regional transport model to interpret $CH_4$ measurements at the Gale Crater which suggested that a large, continuous source is required to satisfy the observations. General Circulation Model simulations (Viscardy et al., 2016) suggested mixing timescales of a few weeks in the low to mid atmosphere which implied that the detection of vertical layering could be evidence of recent emission.

Suggested sinks of Martian $CH_4$ include oxidation via hydrogen peroxide ($H_2O_2$) on the surface (Gough et al., 2011) and heterogeneous reactions on dust grains (Farrell et al., 2006; Jensen et al., 2014; Escamilla-Roa et al., 2018). Photochemical removal is also a potentially significant sink for $CH_4$ in the Martian atmosphere. Korablev et al. (2021) presented the first detection of (1-4ppbv) hydrogen chloride (HCl) in the Martian atmosphere possibly associated with a suggested heterogeneous chlorine source from salt aerosol (Olsen et al., 2021; see also Civiš et al., 2019). Enhanced chlorine abundances could present an additional photochemical sink for $CH_4$ on Mars. Photochemical models of Mars have made considerable headway in recent years, both for the 1D (e.g. Boxe et al., 2014; Stock et al., 2017; Krasnopolsky, 2019; Lo et al., 2020) and the 3D (e.g. Lefèvre, et al., 2004; Daerden et al., 2019) models. Sources of uncertainties in the chemical modules include e.g. the photochemical rate coefficients and (for the column models) the assumed eddy transport profile (see e.g. Michaels and Rafkin, 2004). Established photochemical gas-phase sinks in the Martian atmosphere include reaction with (1) hydroxyl (OH), (2) electronically-excited atomic oxygen ($O(^1D)$), and (3) photolysis. The photochemical removal rate depends on the water abundance and the UVB environment. These in turn affect OH and $O(^1D)$ abundances which can react with $CH_4$ (see e.g. Viscardy et al., 2016). Lefèvre and Forget (2009) (see also Mumma et al., 2009) suggested that $CH_4$ observations could not be accounted for by known physics and chemistry. They calculated a $CH_4$ atmospheric e-folding lifetime, $\tau_{CH4}=[CH_4]/(CH_4 \text{ loss rate})$ of ~200 days as constrained by observations (their study derived this value by fitting an idealized tracer in their general circulation model to reproduce an enhancement in $CH_4$ by a factor 4 to 5 in the assumed emission region of the model cell containing Syrtis Major (10$^o$N, 50$^o$E) for 60 sols compared with the global mean $CH_4$ abundance at vernal equinox), which was over-estimated by a factor of ~600 times in the photochemical models. Some works (e.g. Mumma et al., 2009; Giuranna et al., 2019) suggested this factor could be even higher.

A possible implication of the above discrepancy is that the models could be missing some crucial $CH_4$ removal process(es). High Energy Particles (HEPs) could also impact $CH_4$ on Mars although their effect and in particular the influence of induced ion photochemistry is not well determined. Molina-Cuberos et al. (2002), González-Galindo et al. (2013); Bougher et al. (2015) and Cardnell et al. (2016) applied photochemical ion models which suggested e.g. $CO_2^+$ and $O_2^+$ to be some of the primary positive ions and hydrated ions such as $CO_3^-(H_2O)_n$ to be one of the main negative ions during high energy particle bombardment.

In the present paper we assess the extent to which various phenomena could address the Martian methane atmospheric lifetime discrepancy. First, we apply an atmospheric model where we vary in a Monte Carlo approach chemical reaction rates and eddy mixing coefficients within their current uncertainties. Related to this, we analyze the atmospheric pathways of $CH_4$ removal using our unique pathway analysis program. Second, we implement into our atmospheric model neutral dissociation for a range of gas-phase species by low energy (defined here as ranging from one half to five thousand electron volts (eV)) electrons generated from HEPs. Third, we investigate the photochemical response induced by ion-pair production (IPP) (leading to the generation of e.g. electrons with energies typically in the keV to MeV range) associated with Galactic and Solar Cosmic Rays. We then quantify the influence of these phenomena upon the modelled Martian $CH_4$ photochemical lifetime. Section 2 describes the models used. Section 3 presents the scenarios performed. Section 4 shows the results and section 5 draws brief conclusions.

## 2. Model descriptions

### 2.1 Photochemical Module (BLACKWOLF)

We use the BerLin Atmospheric Chemical Kinetics and photochemistry module With application to exOpLanet Findings (BLACKWOLF) which calculates global mean stationary conditions over the atmospheric column. In the present work the BLACKWOLF chemistry module is adapted to Mars conditions as follows. We take a fixed modern Mars temperature profile input based on Haberle (2017). We employ the standard reference atmosphere for clear (low dust loading) conditions. The temperature ranges from about 216K at the surface, decreasing to about 145K at around 60km and remains approximately at this value in the middle atmosphere. The vertical grid contains 100 equally-spaced model layers extending from the Martian near-surface up to 0.01 Pa. For Mars conditions we employ a surface Bond albedo equal 0.25 (Williams, 2010). The Eddy diffusion coefficients for the Martian atmosphere are calculated as function of temperature and scale height based on Gierasch and Conrath (1985) and are shown in Wunderlich et al. (2020) their Figure 1c. The model lid features a parameterized escape for $O(^3P)$ and H. Details of the full chemical network used our study can be found in Wunderlich et al (2020). The chemical network has 1127 chemical reactions for 128 species, including 832 bimolecular reactions, 117 termolecular reactions, 53 thermal dissociation reactions, and 125 photolysis reactions for 81 absorbers from 100 to 850nm in 133 bands and the same eight Rayleigh absorbers as described in Scheucher et al. (2020). The chemical scheme includes chemical families of hydrogen, nitrogen, carbon, chlorine and sulfur reactions with hydrocarbons up to C5 and features no hazes and no dust. At the surface $CO_2$ and $N_2$ are fixed to a volume mixing ratio (vmr) of 0.9532 and 0.027 respectively whereas $SO_2$ and HCl surface fluxes are fixed to $1.5 \times 10^6$ and $2.4 \times 10^4$ molecules $cm^{-2}\ s^{-1}$ (see Wunderlich et al., 2020, their Table 6 for further details). Modeled chemical output for modern Mars conditions compares reasonably well with earlier model studies such as Nair et al. (1994) and Krasnopolsky (2010) as shown in Wunderlich et al. (2020) their Figure 3. Further details of the photochemical reaction rate coefficient databases can be found in Wunderlich et al. (2020).

### 2.2. Pathway Analysis Program

The Pathway Analysis Program (PAP) (Lehmann, 2004) enables the identification and quantification of chemical production and loss pathways for a given prescribed species in arbitrary chemical systems. The algorithm begins with individual reactions as pathways and continues by connecting shorter pathways at the so-called branching species which are processed according to their lifetime beginning with the short-lived. If a pathway contains sub-pathways (which denote a subset of chemical reactions which belong to the pathway) then it is split into these. The calculation ends if a branching point species with a lifetime greater than a user-defined threshold, $\tau_{max}$, is reached. Alternatively, the algorithm may stop at a user-defined branching point species. To avoid excessive computational times, the algorithm identifies and rejects pathways whose rate falls below a prescribed threshold ($f_{min}$) during the computation. The PAP algorithm takes as input: (1) a list of species names and reactions, (2) chemical reaction rates output by the model averaged over a given time interval (in the present study PAP takes as input the arithmetic mean of rates and abundances from the last two consecutive outputs of the chemistry model before convergence which is used for determining the relative contributions of material flux for the cycles identified) and (3) the concentration change which the chemical reaction rates would produce alone (see also Stock et al., 2012).

In the following analysis the PAP input parameter $f_{min}$ was set to $10^{-16}$ ppbv/s. This parameter is used to avoid so-called "combinatorial explosion" (excessive cpu time required) whereby pathways found with rates below $f_{min}$ are removed from the PAP calculation step-by-step as they are found. Further tests suggested that the value used was appropriate for the system investigated; decreasing $f_{min}$ further did not change the output appreciably. For the pathway analysis it is important to specify a suitable timescale over which the pathways operate. In this study we analyze pathways having branching-point species (c.f. above) with lifetimes less than the photochemical lifetime of carbon monoxide (CO). We choose this species because it is a stable product of the $CH_4$ degradation (about 6 years according to Krasnopolsky, 2007). Note that the pathways for the final conversion of CO to $CO_2$ have been investigated by Stock et al. (2012). For a (hypothetical) chemical model having only in-situ gas-phase reaction rates, the change in chemical concentrations would be accounted for ("chemically balanced") purely by the chemical reaction rates. In the present analysis we only investigate changes of species' concentrations that are caused by chemical reactions. Other processes such as mixing, emissions, escape or deposition could be treated as "pseudo reactions", but this was not done in the present study. Species for which the concentration change was not balanced by reaction rates were: $H_2CO$, $O(^3P)$, $H_2O$, OH, $HO_2$, $H_2O_2$, $O_3$, H, $H_2$, $CH_4$, CO, $CH_3OOH$, $CH_3O_2$, $N_2O$, NO, $NO_2$, $HNO_2$, $HNO_3$, $HO_2NO_2$, $NO_3$, $N_2O_5$, $Cl_2O_2$, $CH_3Cl$, HOCl, Cl, ClO, HCl, $ClONO_2$, $H_2S$, HS, SO, $SO_2$, $H_2SO_4$, HSO, $SO_{4_aerosol}$, $^1CH_2$, $^3CH_2$, $O(^1D)$, $CH_3$, $H_3CO$, HCO, N, NOCl, ClONO, $ClO_2$, $Cl_2$, S, $SO_2^1$, $SO_2^3$, $HSO_3$, $SO_3$, $S_2$, $O_2$, $CO_2$ and $N_2$.

**2.3 Ion-neutral chemistry model (ExoTIC)**

The Exoplanetary Terrestrial Ion Chemistry Model (ExoTIC) is based on the University of Bremen Ion Chemistry Model initially developed for Earth's mesosphere and lower thermosphere (Winkler et al., 2008; Sinnhuber et al., 2012) but generalized to consider any kind of terrestrial (rocky) planet and stellar environment. The time-dependent column model includes 60 neutral and 120 charged species. Neutral and ion photochemistry is driven by input stellar fluxes in the range 120-800nm, considering photodissociation and ionization including particle-impact ionization. Photoelectron detachment and attachment were included for neutral-neutral, ion-neutral and ion-ion reactions. Primary ions are calculated from the excitation, ionization and dissociative ionization of $O_2$, $N_2$, O, $CO_2$ and CO based on their respective abundances and the ionization rate. The vertical domain and input temperature as well as the initial atmospheric composition were the same as those for BLACKWOLF (see 2.1). The effect of particle impact profiles calculated in ExoTIC were provided as input into BLACKWOLF. Further details of ExoTIC can be found in Herbst et al. (2019a).

**2.4 PLANETOCOSMICS and AtRIS**

To model the Martian ionization rates we have two simulation codes at hand: *PLANETOCOSMICS* (Desorgher, 2005), a C++-based GEANT4-driven code which recently has been used model the Martian radiation dose of modern Mars (see, e.g., Gronoff et al., 2015; Köhler et al., 2016 and Matthiä and Berger, 2017) and the Atmospheric Radiation Interaction Simulator (*AtRIS*; Banjac, Herbst, and Heber, 2019), the newly-developed C++-code to model hadronic and electromagnetic particle interactions within a variety of (exo)planetary atmospheres, which has been applied to model modern Earth (Banjac et al., 2020), Mars (see, Guo et al., 2019; Röstel et al., 2020), Venus (Herbst et al., 2019a; 2020), and Earth-like exoplanets (Herbst et al., 2019b; Scheucher et al., 2020).

In a first step, we utilize both codes to calculate the Ion Pair Production (IPP) profiles in the Martian atmosphere which arise due to the impact of Galactic Cosmic Rays (GCRs, see Section 3.6), while in a second step also Solar Proton Events (SPEs) are taken into account (see Section 3.7). For both scenarios, the model inputs include information on the cosmic ray fluxes on top of the atmosphere (here based on the LIS by Herbst et al., 2017) and the magnetic field (e.g., Herbst et al., 2013). Note that in our study the influence of the crustal magnetic field bubbles is neglected.

**3. Scenarios**

**3.1 Mars Control Run with BLACKWOLF: Gas Phase Chemistry only, without HEPs**

The control run featured gas-phase chemistry only i.e. without high energy particle reactions from HEPs. The surface $CH_4$ value was fixed to the 38-month mean $CH_4$=0.41 ppbv observed at Gale Crater by Curiosity (Webster et al., 2018). The surface $CH_4$ flux in BLACKWOLF which would be required to maintain this concentration against the atmospheric chemical sinks corresponds to 9.16 Mg/yr. This value is about 61 million times weaker than on modern Earth which has ~558 Tg/year in the global mean (see e.g. Saunois et al., 2016). Earlier budget estimates (e.g. Summers et al., 2002) suggested a $CH_4$ surface flux weaker by at least a factor 100,000 on Mars compared with Earth. Further details of the Mars control model setup and boundary conditions can be found in Wunderlich et al. (2020).

**3.2 Monte Carlo Simulations of the Control Run**

**3.2.1 Varying chemical rate coefficients and eddy coefficients within their uncertainty range**

The BLACKWOLF photochemical module was run with the same basic setup as scenario 3.1 but now in a Monte Carlo (MC) setup whereby key model parameters were varied randomly within their uncertainty range. A subset of chemical rate coefficients important to simulate the atmospheric composition of modern Mars taken from Krasnopolsky (2019) (his chapter 10) (see below and Appendix 1a, 1b) were varied. This conservative approach ensured that our output range of species abundances was broadly kept within the current observational constraints of modern Mars (but see discussion below). For the three-body reactions, current rate coefficients in BLACKWOLF are based on observations in an $N_2$-$O_2$ bath gas but are expected to have higher values on moving to a $CO_2$ bathgas (see e.g. Nair, 1994; Krasnopolsky, 2019) since the triatomic $CO_2$ molecule can accommodate a wider range of energy states than diatomic $O_2$ and $N_2$. Note that Lindner (1988) discussed termolecular reaction rate coefficients for Mars conditions. The termolecular rate coefficients were therefore varied randomly up to the limit shown for the reactions presented in Appendix 1a. These limits are based on the observed increase in termolecular rate coefficients for a nitrogen compared with a carbon dioxide bath gas taken from Nagy et al. (2015) (their Table 3). All photolysis rates ($P_r$) in BLACKWOLF were varied within a factor two i.e. from $P_{r(min)}$ to $P_{r(max)}$, where $P_{r(max)}/P_{r(min)} = 2$. The eddy mixing coefficients in the model were varied randomly and independently of each other at the surface between ($10^4$-$10^6$) $cm^2\ s^{-1}$ and at the Top of Atmosphere (TOA) between ($10^6$-$10^8$) $cm^2\ s^{-1}$ (see e.g. Nair, 1994; Krasnopolsky, 2019). A logarithmic interpolation was employed between the surface layer and the TOA.

The MC runs originally featured N=2200 individual model simulations (hereafter referred to as realizations) of which 1000 were shown in the final Figures and used in the resulting analysis since these remained within the observational range for CO and $O_2$ (see below). As a test we performed an additional simulation with N=5000. Results suggested that the N=1000 case was sufficient for our purposes since the range of uncertainties calculated for the various species abundances changed only by up to a few percent compared with N=5000. The MC realizations all ran with the full BLACKWOLF chemical network. For the bimolecular gas-phase reactions in the varied subset the uncertainty in the reaction rate constant as a function of temperature (f(T)) was taken from the Jet Propulsion Laboratory (JPL) Data Evaluation Report 19 described in Burkholder et al. (2020) (see Appendix 1b) and is equal to:

$$f(T) = f(298\ K)\ \exp|g\cdot(1/T - 1/298\ K)| \qquad (1)$$

where g is constant for a given reaction. The error interval from (1) is obtained by multiplying (upper error limit) and dividing (lower error limit) the rate coefficient by $f^2(T)$ corresponding to the 95% confidence interval (Burkholder et al., 2020). Therefore, for the MC analysis we multiplied each bimolecular rate constant by its individual uncertainty factor (U) where:

$$U = r \cdot f(T)^2 + (1 - r) \cdot 1/f(T)^2 \qquad (2)$$

where r is a random number between (0-1), f(T) is taken from Appendix 1a and was calculated for the Martian temperature profile. Note the caveat that equation (2) assumes that 100% of the data lie in the interval $[1/f(T)^2, f(T)^2]$ which corresponds to the 95% confidence interval discussed by the Burkholder study. On the other hand, equation (2) provides a straightforward means of estimating U given that neither f nor g are derived from a rigorous statistical treatment. Note that the MC analysis takes as input the reported range of uncertainty in the chemical rate coefficients and the Eddy mixing and then calculates the associated variation range in the species abundances.

### 3.2.2 Varying the eddy coefficients only

In order to separate the effect of the chemical rate uncertainty from that of the eddy coefficients, scenario 3.2.2 is as for 3.2.1 but varied the eddy coefficients only.

### 3.3 Variation of atmospheric water profile

The effect of changing the input water profile vmr was investigated by fixing the $H_2O$ vmr to isoprofiles with vmr values of $10^{-4}$, $3x10^{-4}$ and $10^{-3}$ for three different regions in the lower atmosphere i.e. in the surface layer (0-1km) only; (1-10km) or (1-20km) (see Figure 8, upper left panel). A logarithmic interpolation was employed between $h_c$ and 50 km, where $h_c$ denotes the level at the top of the height range where $H_2O$ vmr is set to a constant isoprofile. At 50 km and above the $H_2O$ was calculated interactively by BLACKWOLF (see Figure 8). This scenario also ran with the full BLACKWOLF network and with N=1000 MC simulations as for 3.2. Trokhimovskiy et al. (2015) and Fedorova et al. (2020) discuss the Martian water cycle. Note that recent near global observations of the seasonal variation in the Martian water profile with the ExoMars TGO (Villanueva et al., 2021, their Figure 2, left panels) suggests (0.8 – 1.2) $x10^{-4}$ vmr $H_2O$ in the lower atmosphere decreasing to around $4x10^{-5}$ from ~50 to 80 km which are somewhat lower vmr values than assumed in our work (see Figure 7, top left panel).

**3.4 Variation of atmospheric chlorine burden**

In light of the recently detected HCl (Korablev et al., 2021) we performed a run (scenario 3.4) as for run 3.1 (which featured ~0.2 ppbv HCl at the surface based on the upper limit of Hartogh et al., 2010) but with enhanced chlorine. Two cases were studied featuring firstly, surface HCl set to 2 ppbv and with no surface source of chloromethane ($CH_3Cl$) (case 3.4.1) and secondly, with surface HCl set to 2 ppbv and surface $CH_3Cl$ set to 14 ppbv based on the upper limit of Villanueva et al. (2013) (case 3.4.2). Note that recent observations of HCl (Korablev et al., 2021; Aoki et al., 2021) suggested variability in time and space with values of (1-3) ppbv from (10-40km) at perihelion decreasing to around a few tenths of a part per billion above ~30 km during the solstice. These works suggest that HCl is quickly disappearing at the close of the southern summer, more rapidly than its simulated photochemical lifetime. More work is required to understand the seasonal responses on a global scale. Note too that since recent observations of HCl (see above) suggest values of a few ppbv, further work is required to investigate if the assumed upper limit of 14 ppbv $CH_3Cl$ in our work is consistent with the HCl data.

**3.5 As for control run (3.1) but with low energy electron dissociation of gas-phase species for flaring conditions**

On Earth secondary particles induced from air shower events require energies typically ranging from KeV to MeV to reach the surface (see e.g. Risse and Heck, 2004). Although the lower energy (~1-1000 eV) secondary electrons feature large collisional cross sections for the neutral dissociation and ionization of atmospheric species, they are absorbed (on Earth) mainly above the mesosphere and likely have a negligible effect upon the middle atmosphere composition. However, in the thin Martian atmosphere, their depth of penetration and influence upon atmospheric species is not well-determined. Run 3.5 includes the neutral dissociation for a range of Martian atmospheric species. Appendix 2 shows the species included, their cross sections used, and describes the derivation of flux profiles on Mars during flaring conditions for two different methods.

**3.6 Modern Mars for solar minimum conditions including GCRs**

This scenario was as for 3.1 but additionally assumed quiescent (solar minimum) conditions with the GCR TOA input proton energy spectrum based on the LIS by Herbst et al. (2017, see black line in Figure 1). Simulations are performed for both PLANETOCSOMICS and AtRIS, a comparison of the results is discussed in Section 4.7.1.

**3.7 Modern Mars for SPE flaring conditions**

This scenario was as for 3.1 but additionally assumed flaring conditions of a) the February 1956 Ground Level Enhancement (GLE05, Raukunen et al., 2018), the strongest GLE event measured directly so far, b) the Carrington event based, and c) the AD774/775 event, one of the strongest GLEs ever detected in the cosmogenic radionuclide records of $^{10}Be$, $^{14}C$, and $^{36}Cl$ (e.g., Brehm et al., 2021, and references therein). Information on the actual event spectra of the Carrington event and AD774/775, however, are missing. Typically, the events are scaled based on known GLE events. In this study, the Carrington event was scaled based on GLE44 (October 1989) with a rather flat spectrum but much higher intensities in the low energy part, while AD774/775 usually is scaled with the hard GLE05 spectrum. According to Usoskin and Kovaltsov (2021) the AD774/775 event can be assumed to be up to 70 times stronger than GLE05. In addition, the solar particle flux strongly depends on the orbital distance. Thus, in all three cases we applied a $1/R^2$ scaling from Earth to Mars (see colored lines in Figure 1). The results are discussed in Section 4.7.1.

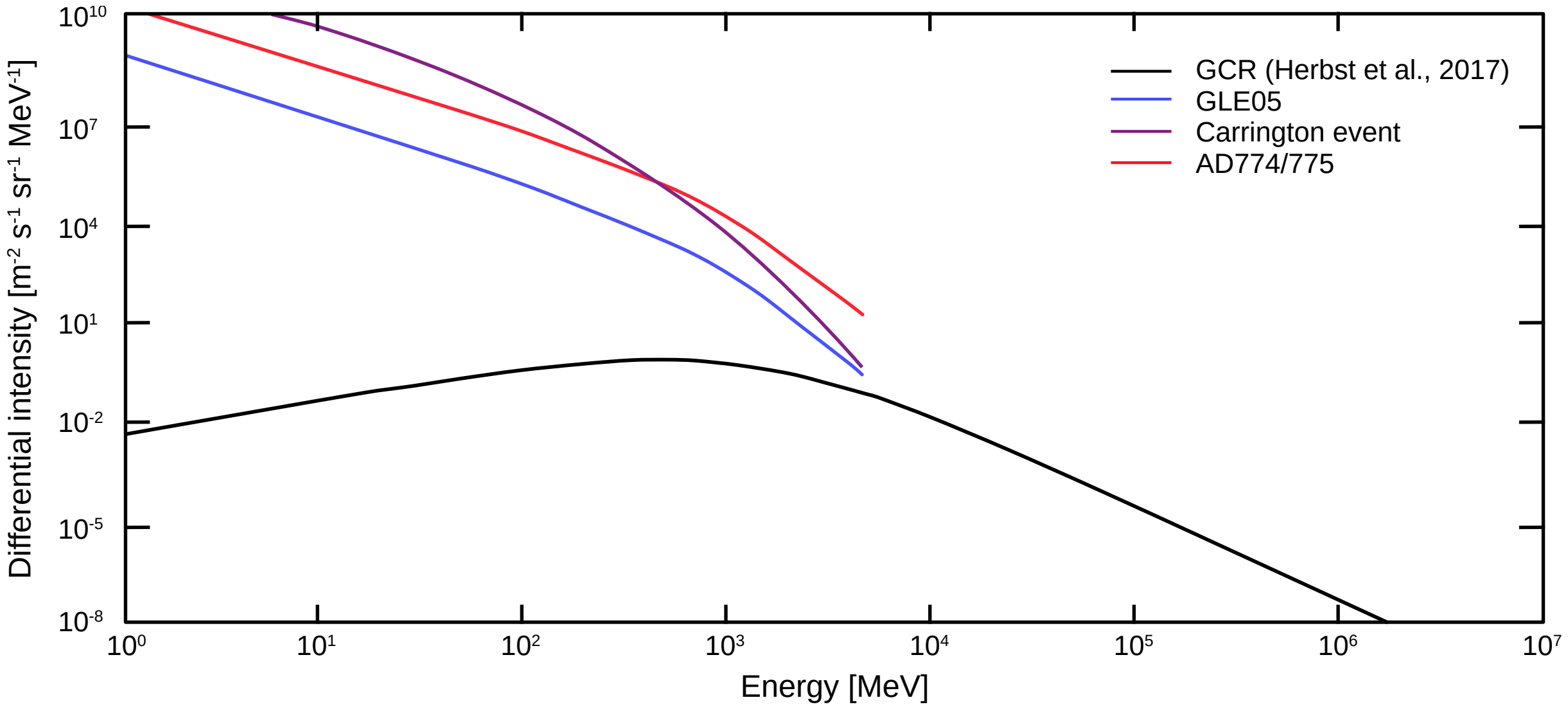

*Figure 1: Proton spectra of GCRs (black line) and the three strong GLE events studied: GLE05 (blue line), Carrington event (purple line), and AD774/775 (red line).*

**4. Results**

**4.1 Mars Control Run (Scenario 3.1)**

Figure 2 shows the rates of $CH_4$ loss reactions (molecules cm$^{-3}$ s$^{-1}$) based on the computations with BLACKWOLF for our Mars control run (scenario 3.1):

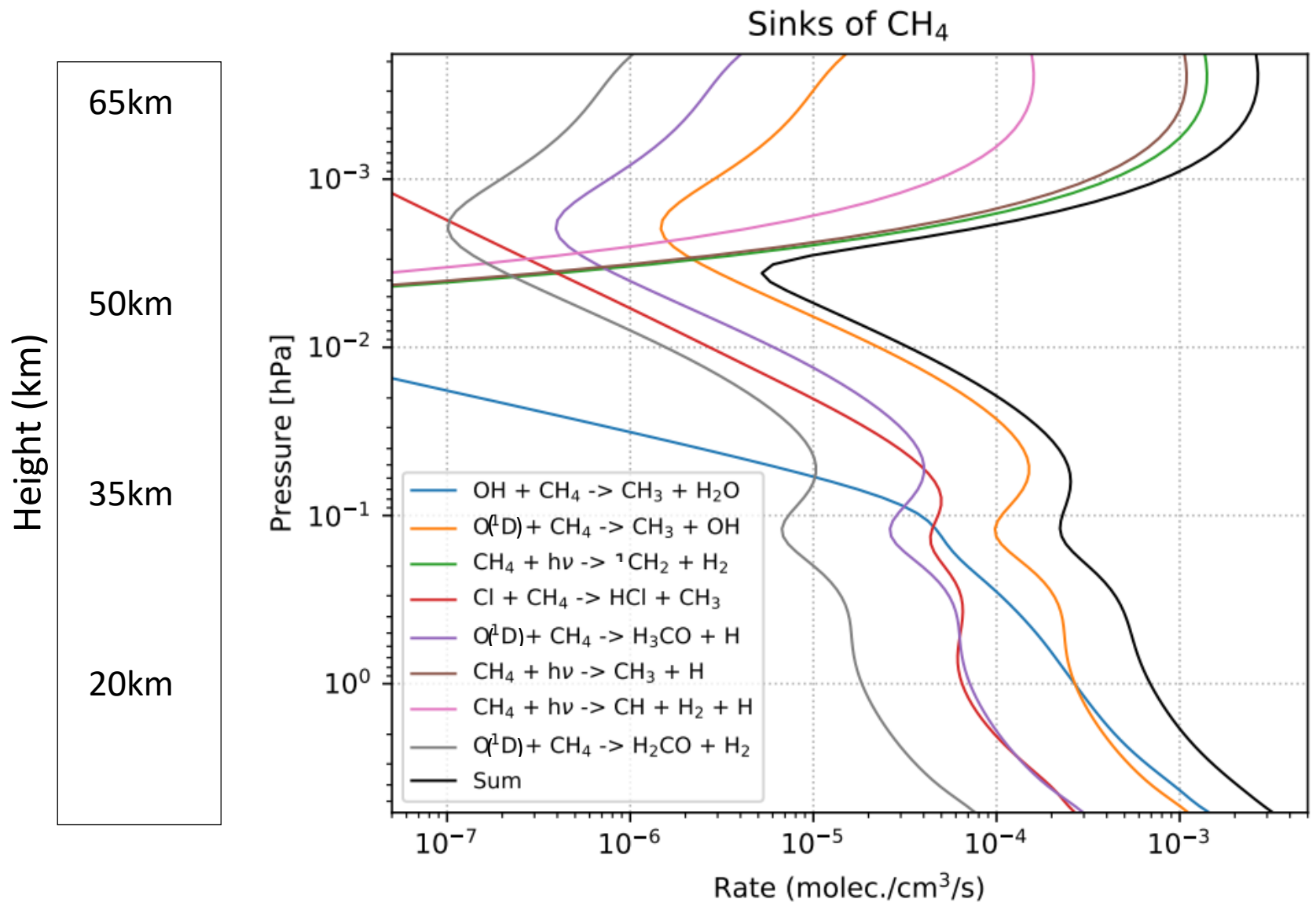

*Figure 2: Profiles showing $CH_4$ loss rates (molecules cm$^{-3}$ s$^{-1}$) calculated by the BLACKWOLF model for the Mars control run (scenario 3.1).*

Figure 2 suggests that the $CH_4$ loss occurs mainly by reactions with OH and O($^1$D) in the lower model layers, with O($^1$D) in the middle layers and by photolysis in the uppermost layers.

Figure 3 shows the contribution of atmospheric $CH_4$ loss initiated by reaction with OH, Cl, O($^1$D) and photolysis in the Mars control scenario (3.1):

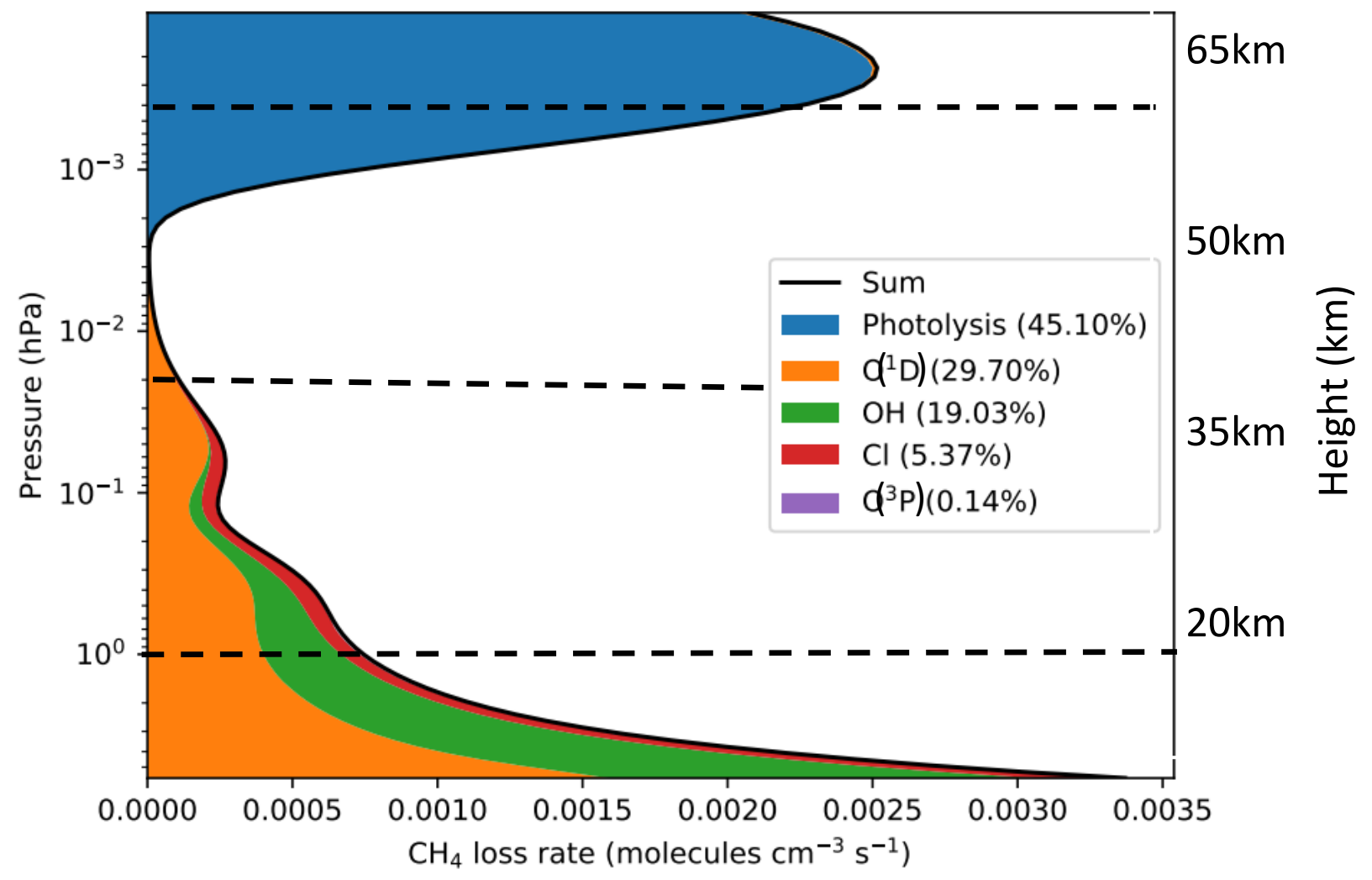

*Figure 3: Photochemical loss (molecules cm$^{-3}$ s$^{-1}$) of $CH_4$ in the Mars control scenario (3.1). Black dashed lines indicate the three pressure levels investigated in Table 1.*

Values in the legend show the column integrated percentage $CH_4$ loss. Whereas Figure 2 shows the rates of individual chemical reactions, Figure 3 sums together the rates of all reactions indicated in the legend. For example "photolysis" in the legend of Figure 3 indicates the column-integrated sum of all reaction rates where $CH_4$ destruction is initiated via photolysis (e.g. $CH_4$+hν→$CH_3$+H, $CH_4$+hν→CH+$H_2$+H…) and similarly for the reaction of O($^1$D) with $CH_4$ etc. Figure 3 suggests that the main $CH_4$ removal reactions over pressure are comparable to those discussed for Figure 1. An important source of O($^1$D) in the model was photolysis of $O_3$ whereas an important source of OH was the photolysis of $H_2O$. Figure 3 shows only the first reaction step in the $CH_4$ atmospheric degradation. In the following section we discuss the full degradation pathways. Compared with Figure 3, model results from Krasnopolsky et al. (2004) (their Table 1) suggested that 63% of their column-averaged atmospheric $CH_4$ was removed via photolysis, ~31% via O($^1$D) and ~6% via OH. Our work therefore suggests that photolytic removal is somewhat less important whereas removal via OH is somewhat more important than the Krasnopolsky study, attributable e.g. to our updated rate and photolytic coefficients taken e.g. from the HITRAN (2016) database, different assumptions in the $H_2O$ profile and our more extended chemical network. Our results are also consistent with model estimations by Lefèvre (2019) which suggested that OH and O($^1$D) together represent 40-50% of the total loss of $CH_4$ depending on season. Figure 3 shows only the $CH_4$ *loss* reactions in the Mars control run (note that $CH_4$ *production* was essentially zero (not shown) since the largest amount of $CH_4$ by far enters the atmosphere as a surface source with very minor (<<1%) contributions from atmospheric in-situ sources via e.g. organic degradation reactions).

**4.2 Pathway Analysis of Mars Control Run**

Atmospheric $CH_4$ degradation on Earth can involve NOx catalysis (see also text to Table 1). On Mars, however the NOx abundance is reduced. In our Mars control run, for example, the surface NOx (=NO+$NO_2$) is 0.18 ppbv (NO=0.10 ppbv, $NO_2$=0.08 ppbv). These values agree well with the 3D model study of Moudden and McConnell (2007) who reported NO=0.1 ppbv at the surface (their Figure 6b). Krasnopolsky (2006) suggested an upper limit of NO=1.7 ppbv in the Martian lower atmosphere. Due to the different ambient conditions on Mars compared to Earth, different pathways of the oxidation of $CH_4$ to $CO_2$ can be expected, which are less well known. We therefore apply the PAP tool to study the $CH_4$ degradation in the Martian atmosphere in more detail.

**Pathway Analysis Results**

Table 1 shows the five most important pathways, as determined by PAP, for three atmospheric levels which are referred to as the Upper Regime (UR at $5x10^{-4}$ hPa), the Middle Regime (UR at 0.02 hPa) and the Lower Regime (LR at 1hPa):

Table 1: The five most important pathways on the Upper Regime (UR at $5x10^{-4}$ hPa), Middle Regime (MR at 0.02 hPa) and Lower Regime (LR at 1hPa) as marked by horizontal black dashed lines in Figure 3. Values show the net $CH_4$ loss (in molecules $cm^{-3}$ $s^{-1}$) in the Martian atmosphere as a % of the total loss found by PAP on the level considered. For a given pathway the bottom reaction written below the horizontal dashed black line denotes the overall net reaction. 'M' denotes any chemical species required to remove excessive vibrational energy.

| | | | | | |
|---|---|---|---|---|---|
| **UPPER** | **Pathway UR1: 17.71%**<br>$CH_4 + h\nu \rightarrow {}^1CH_2 + H_2$<br>${}^1CH_2 + M \rightarrow {}^3CH_2 + M$<br>$O(^3P) + {}^3CH_2 \rightarrow CO + H + H$<br>----------------------------------<br>$CH_4 + O(^3P) \rightarrow CO + H_2 + 2H$ | **Pathway UR2: 11.84%**<br>$CH_4 + h\nu \rightarrow CH_3 + H$<br>$O(^3P) + CH_3 \rightarrow CO + H_2 + H$<br>----------------------------------<br>$CH_4+O(^3P) \rightarrow CO + H_2 + 2H$ | **Pathway UR3: 10.77%**<br>$CH_4 + h\nu \rightarrow {}^1CH_2 + H_2$<br>${}^1CH_2 + M \rightarrow {}^3CH_2 + M$<br>$O(^3P) + {}^3CH_2 \rightarrow CO + H_2$<br>--------------------------------<br>$CH_4 + O(^3P) \rightarrow CO + 2H_2$ | **Pathway UR4: 9.36%**<br>$CH_4 + h\nu \rightarrow CH_3 + H$<br>$O(^3P) + CH_3 \rightarrow H_2CO + H$<br>$H_2CO + h\nu \rightarrow H_2 + CO$<br>--------------------------------<br>$CH_4+ O(^3P) \rightarrow CO + H_2 + 2H$ | **Pathway UR5: 5.71%**<br>$CH_4 + h\nu \rightarrow CH_3 + H$<br>$O(^3P) + CH_3 \rightarrow H_2CO + H$<br>$H_2CO + h\nu \rightarrow HCO + H$<br>$O(^3P) + HCO \rightarrow CO_2 + H$<br>--------------------------------<br>$CH_4 + 2O(^3P) \rightarrow 4H + CO_2$ |
| **MIDDLE** | **Pathway MR1: 9.94%**<br>$O(^1D)+CH_4 \rightarrow CH_3 + OH$<br>$O(^3P)+CH_3 \rightarrow CO+H_2 + H$<br>$O(^3P) + OH \rightarrow O_2 + H$<br>$3[ CO_2+h\nu \rightarrow CO+ O(^3P)]$<br>$O(^3P) +O_2+M \rightarrow O_3 + M$<br>$O_3 + h\nu \rightarrow O_2 + O(^1D)$<br>-------------------------------<br>$CH_4+3CO_2 \rightarrow O_2+4CO+H_2+2H$ | **Pathway MR2: 7.32%**<br>$O(^1D) + CH_4 \rightarrow CH_3 + OH$<br>$O(^3P) + CH_3 \rightarrow H_2CO + H$<br>$O(^3P)+H_2CO \rightarrow HCO + OH$<br>$HCO + O_2 \rightarrow CO + HO_2$<br>$O(^3P) + HO_2 \rightarrow OH + O_2$<br>$3[O(^3P) + OH \rightarrow O_2 + H]$<br>$7[CO_2+ h\nu \rightarrow CO + O(^3P)]$<br>$O(^3P) + O_2 + M \rightarrow O_3 + M$<br>$O_3 + h\nu \rightarrow O_2 + O(^1D)$<br>---------------------------------<br>$CH_4+7CO_2 \rightarrow 3O_2+8CO+4H$ | **Pathway MR3: 6.41%**<br>$O(^1D) + CH_4 \rightarrow CH_3 + OH$<br>$O(^3P) + CH_3 \rightarrow CO + H_2 + H$<br>$O(^3P) + OH \rightarrow O_2 + H$<br>$O(^3P) + O_2 + M \rightarrow O_3 + M$<br>$O_3 + h\nu \rightarrow O_2 + O(^1D)$<br>------------------------------------<br>$CH_4+3O(^3P) \rightarrow O_2+CO+ H_2+2H$ | **Pathway MR4: 6.08%**<br>$O(^1D) + CH_4 \rightarrow CH_3 + OH$<br>$O(^3P) + CH_3 \rightarrow H_2CO + H$<br>$H_2CO + h\nu \rightarrow HCO + H$<br>$HCO + O_2 \rightarrow CO + HO_2$<br>$O(^3P) + HO_2 \rightarrow OH + O_2$<br>$2[O(^3P) + OH \rightarrow O_2 + H]$<br>$5[CO_2 + h\nu \rightarrow CO + O(^3P)]$<br>$O(^3P) + O_2 + M \rightarrow O_3 + M$<br>$O_3 + h\nu \rightarrow O_2 + O(^1D)$<br>-----------------------------------<br>$CH_4+5CO_2 \rightarrow 2O_2+ 6CO+ 4H$ | **Pathway MR5: 5.50%**<br>$O(^1D) + CH_4 \rightarrow CH_3 + OH$<br>$O(^3P) + CH_3 \rightarrow H_2CO + H$<br>$H_2CO + h\nu \rightarrow H_2 + CO$<br>$O(^3P) + OH \rightarrow O_2 + H$<br>$3[ CO_2 + h\nu \rightarrow CO + O(^3P)]$<br>$O(^3P) + O_2 + M \rightarrow O_3 + M$<br>$O_3 + h\nu \rightarrow O_2 + O(^1D)$<br>--------------------------------------<br>$CH_4+3CO_2 \rightarrow O_2+4CO+ H_2 + 2H$ |
| **LOWER** | **Pathway LR1: 6.99%**<br>$OH + CH_4 \rightarrow CH_3 + H_2O$<br>$CH_3 + O_2 \rightarrow H_2CO + OH$<br>$H_2CO + h\nu \rightarrow H_2 + CO$<br>-------------------------------<br>$CH_4+O_2 \rightarrow CO+ H_2 + H_2O$ | **Pathway LR2: 6.41%**<br>$OH + CH_4 \rightarrow CH_3 + H_2O$<br>$CH_3+O_2+M \rightarrow CH_3O_2 + M$<br>$CH_3O_2+h\nu \rightarrow H_2CO + OH$<br>$H_2CO + h\nu \rightarrow H_2 + CO$<br>--------------------------------<br>$CH_4 + O_2 \rightarrow CO + H_2 + H_2O$ | **Pathway LR3: 5.07%**<br>$O(^1D) + CH_4 \rightarrow CH_3 + OH$<br>$CH_3 + O_2 \rightarrow H_2CO+ OH$<br>$H_2CO+h\nu \rightarrow HCO+ H$<br>$HCO + O_2 \rightarrow CO + HO_2$<br>$H + O_2 + M \rightarrow HO_2 + M$<br>$2[ OH + HO_2 \rightarrow H_2O + O_2]$<br>$CO_2 + h\nu \rightarrow CO + O(^3P)$<br>$O(^3P) + O_2 + M \rightarrow O_3 + M$<br>$O_3 + h\nu \rightarrow O_2 + O(^1D)$<br>---------------------------------<br>$CH_4+O_2+CO_2 \rightarrow 2CO + 2H_2O$ | **Pathway LR4: 4.63%**<br>$O(^1D) + CH_4 \rightarrow CH_3 + OH$<br>$CH_3 + O_2 + M \rightarrow CH_3O_2 + M$<br>$CH_3O_2 + h\nu \rightarrow H_2CO + OH$<br>$H_2CO + h\nu \rightarrow HCO + H$<br>$HCO + O_2 \rightarrow CO + HO_2$<br>$H + O_2 + M \rightarrow HO_2 + M$<br>$2[OH + HO_2 \rightarrow H_2O + O_2]$<br>$CO_2 + h\nu \rightarrow CO + O(^3P)$<br>$O(^3P) + O_2 + M \rightarrow O_3 + M$<br>$O_3 + h\nu \rightarrow O_2 + O(^1D)$<br>-----------------------------------<br>$CH_4+ O_2 +CO_2 \rightarrow 2CO+ 2H_2O$ | **Pathway LR5: 4.01%**<br>$OH + CH_4 \rightarrow CH_3 + H_2O$<br>$CH_3 + O_2 \rightarrow H_2CO + OH$<br>$H_2CO + h\nu \rightarrow HCO + H$<br>$HCO + O_2 \rightarrow CO + HO_2$<br>$H + O_2 + M \rightarrow HO_2 + M$<br>$HO_2 + HO_2 \rightarrow H_2O_2 + O_2$<br>-----------------------------------<br>$CH_4+2O_2 \rightarrow CO + H_2O + H_2O_2$ |

The upper regime in Table 1 features pathways involving $CH_4$ photolysis to form ${}^1CH_2$ or $CH_3$. These radical species can react further with $O(^3P)$ to form CO and atomic hydrogen. There is a net consumption of $O(^3P)$ and a net production of H and $H_2$ in the upper regime. In our stationary model these processes are balanced by transport and escape. In our model we calculate escape fluxes of H, $H_2$ and $O(^3P)$ in the upper boundary based on Nair (1994). The e-folding lifetime of modelled H was 3.7 Earth days at $10^{-3}$ hPa (~70 km). Assuming vertical transport rates of a few tens of km in ~5 Earth days (see Holmes et al., 2017, their Figure 2) suggests that the H produced by the upper region pathways in Table 1 could reach the upper model lid (~90 km) within (1-2) e-folding lifetimes and escape. The middle regime in Table 1 features pathways involving reaction of $CH_4$ with $O(^1D)$ whereas the lower regime features pathways involving reaction of $CH_4$ with OH and $O(^1D)$ consistent with

Figure 3. Numerous pathways in Table 1 involve production of CO, H and $H_2$ via net $CH_4$ destruction. Such pathways do not likely influence the abundance of CO, H and $H_2$ due to the low abundance of $CH_4$. Our pathway analysis results suggest that $CH_4$ is unlikely to influence the abundances of CO, H and $H_2$. Related to this, note that there are only $CH_4$ loss reactions and no in-situ $CH_4$ production reactions in the system.

Pathways involving chlorine (which makes up about 5% of the total $CH_4$ removal rate, see Figure 3) did not feature in the main pathways in Table 1 which suggests that the chlorine loss is made up of a large number of pathways individually contributing small (<1%) rates. In the lower region (1hPa) the most important pathway involving chlorine for the control run (3.1) recycled hydrochloric acid (HCl) and contributed 0.66% of the total $CH_4$ loss found by PAP on this level:

$$Cl + CH_4 \rightarrow HCl + CH_3$$
$$CH_3 + O_2 \rightarrow H_2CO + OH + M$$
$$H_2CO + h\nu \rightarrow H_2 + CO$$
$$OH + HCl \rightarrow H_2O + Cl$$
------------------------------------
$$CH_4 + O_2 \rightarrow CO + H_2 + H_2O$$

It is informative at this point to compare how $CH_4$ is degraded in the atmospheres of Mars and Earth. In Earth's stratosphere, the first step of $CH_4$ oxidation is mostly the reaction with OH (as in Earth's troposphere) with a smaller contribution from $O(^1D)$ and a minor contribution from Cl (see e.g., Burnett and Burnett, 1995). However, in several pathways in Table 1, the product $H_2CO$ can nevertheless be formed without NOx e.g. via:

$$CH_3 + O(^3P) \rightarrow H_2CO + H$$
$$CH_3O_2 + h\nu \rightarrow H_2CO + OH$$

In Earth's troposphere $CH_4$ is oxidized by so-called smog cycles which are mostly initiated by the reaction with OH, involve NOx (=NO+$NO_2$) catalysis and can generate $O_3$ via for example the pathway:

$$CH_4 + OH \rightarrow CH_3 + H_2O$$
$$CH_3 + O_2 + M \rightarrow CH_3O_2 + M$$
$$CH_3O_2 + NO \rightarrow NO_2 + CH_3O$$
$$CH_3O + O_2 \rightarrow HO_2 + H_2CO$$
$$HO_2 + NO \rightarrow NO_2 + OH$$
$$2[NO_2 + h\nu \rightarrow O(^3P) + NO]$$
$$\underline{2[O(^3P) + O_2 + M \rightarrow O_3 + M]}$$
$$CH_4 + 4O_2 \rightarrow H_2CO + 2O_3 + H_2O$$

For low NOx abundances in Earth's troposphere, the following pathway, which does not lead to $O_3$ production can feature:

$$CH_4 + OH \rightarrow CH_3 + H_2O$$
$$CH_3 + O_2 + M \rightarrow CH_3O_2 + M$$
$$CH_3O_2 + HO_2 + M \rightarrow CH_3O_2H + O_2 + M$$
$$CH_3O_2H + h\nu \rightarrow CH_3O + OH$$
$$\underline{CH_3O + O_2 \rightarrow HO_2 + H_2CO}$$
$$CH_4 + O_2 \rightarrow H_2CO + H_2O$$

The $H_2CO$ formed in the net reaction can then participate in further smog cycles where it is ultimately oxidized into $CO_2$ and $H_2O$. Note that in the thin Martian atmosphere, the pressure-dependent reaction: $CH_3O_2 + HO_2 + M \rightarrow CH_3O_2H + O_2 + M$ is less favored than on Earth. This is consistent with the result that the lower region pathways in Table 1 do not feature this reaction; it is instead substituted by the photolysis of $CH_3O_2$. The reaction: $CH_3 + O(^3P)$ can also lead to $H_2CO$ formation in the middle and upper regions as shown in Table 1.

On Mars, the NOx abundance is much smaller than on Earth. In the lower modelled region (1hPa) the most important modelled pathway involving NOx-catalyzed $CH_4$ oxidation as is the case for the tropospheric “smog” cycles on Earth featured only a 0.013% contribution to the loss on this level and led to net production of hydrogen peroxide ($H_2O_2$):

$$O(^1D) + CH_4 \rightarrow CH_3 + OH$$
$$CH_3 + O_2 \rightarrow H_2CO + OH$$
$$H_2CO + h\nu \rightarrow HCO + H$$
$$HCO + O_2 \rightarrow CO + HO_2$$
$$3[OH + CO \rightarrow CO_2 + H]$$
$$4[H + O_2 + M \rightarrow HO_2 + M]$$
$$2[HO_2 + HO_2 \rightarrow H_2O_2 + O_2]$$
$$HO_2 + NO \rightarrow OH + NO_2$$
$$NO_2 + h\nu \rightarrow NO + O(^3P)$$
$$O(^3P) + O_2 + M \rightarrow O_3 + M$$
$$O_3 + h\nu \rightarrow O_2 + O(^1D)$$

---------------------------------------

$$CH_4 + 4O_2 + 2CO \rightarrow 3CO_2 + 2H_2O_2$$

These results suggest that the role of NOx with respect to $CH_4$ oxidation is different on Mars compared with Earth. On Earth, NOx takes part in the direct degradation step $CH_3O_2$+NO$\rightarrow$$NO_2$+$CH_3O$ whereas on both planets NO reacts with $HO_2$ and the resulting $NO_2$ undergoes photolysis. $NO_2$ photolysis leads to $O_3$ formation and $O_3$ photolysis yields $O(^1D)$, which reacts with $CH_4$.

### 4.3 Monte Carlo Variation (Scenarios 3.2.1 and 3.2.2)

Figure 4a shows results comparing the control run (scenario 3.1, solid blue line) with the MC runs (scenario 3.2.1, variation of both photochemical rate and eddy coefficients, grey lines):

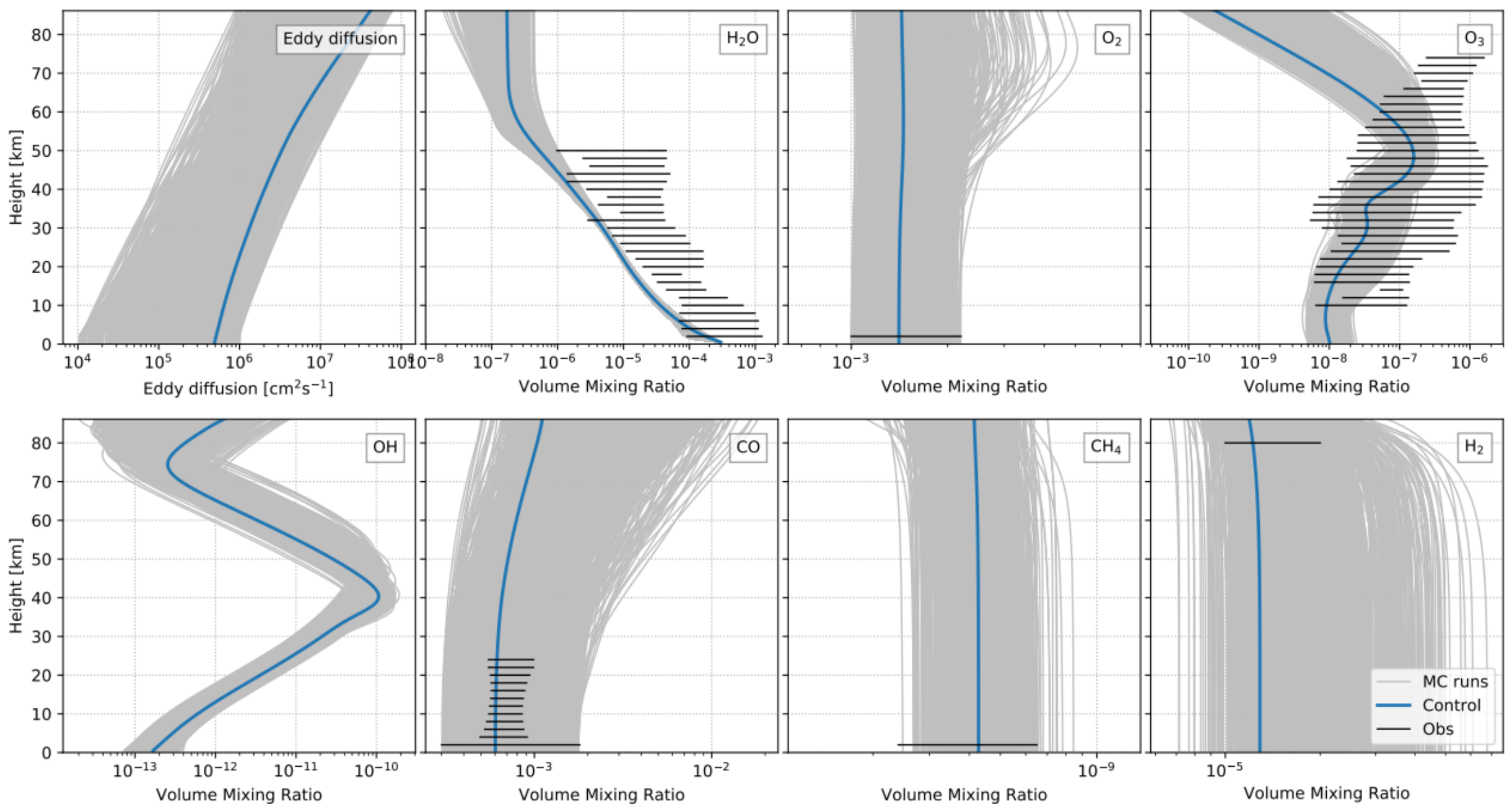


*Figure 4a: Model Profiles for the control run (scenario 3.1) (solid blue lines) and the MC runs (scenario 3.2.1) which vary both photochemical rate and eddy coefficients (solid grey lines for N=1000 realizations, see text) output by BLACKWOLF. Horizontal solid black lines denote the range of observations (see Wunderlich et al., 2020). Species amounts are shown in vmr. The upper panels from left to right show: eddy diffusion coefficient (cm$^2$ s$^{-1}$), $H_2O$, $O_2$ and $O_3$ respectively. The lower panels from left to right show: OH, CO, $CH_4$ and $H_2$ respectively.*

Figure 4a suggests that the control run calculated with the BLACKWOLF model (solid blue line) fits the observational range (horizontal solid black lines, Wunderlich et al., 2020) reasonably well. For $O_3$ (upper right panel in Figure 4a) in the upper layers our model value decreased with altitude whereas the observed value increased. As in our model, Krasnopolsky (2010) (his Figure 10.1) also calculated a decrease in his modelled $O_3$ vmr e.g. by about an order of magnitude from 57km to 75km on Mars. In our model, $O_3$ abundances in the upper layers are mostly determined by the balance between the source reaction: $O+O_2+M \rightarrow O_3+M$ and the sink reaction $O_3+h\nu \rightarrow O_2+O(^1D)$. The difference between model and observations in Figure 4a therefore suggests either that more work is required to investigate the parameterized rates of these two reactions in the cold, thin $CO_2$-dominated upper atmosphere or, that other, unidentified processes (such as transport of $O(^3P)$, see Stock et al., 2017) could be leading to the observed $O_3$ increase with altitude. We reduced the modelled $H_2O$ abundances in Figure 4a in order that CO better compared with the observed values. Using more enhanced $H_2O$ values resulted in enhanced production of OH which led to the modelled CO values (which are generally more accurately known than the $H_2O$ values) lying outside the range suggested by the observations.

The MC scenarios suggests that the final set of N=1000 realizations (solid grey lines) shown in Figure 4a leads to a variation in species concentrations and eddy diffusion coefficient by a factor of up to ten and one hundred respectively depending on species and altitude. From the original MC realizations we discarded those results which fell outside the observed global mean abundance range for the more-accurately known species CO and $O_2$. The CO observations are based on vertical

retrieval via the PFS on Mars Express (Bouche et al., 2019). Here, the horizontal black lines represent uncertainty in the CO retrieval. The filtering of the MC results was weighted more towards the lowermost level in Figure 4a since the observations have enhanced sensitivity in this region. Figure 4b is as for Figure 4a but shows the MC realizations for scenario 3.2.2 which varies only the eddy coefficients:

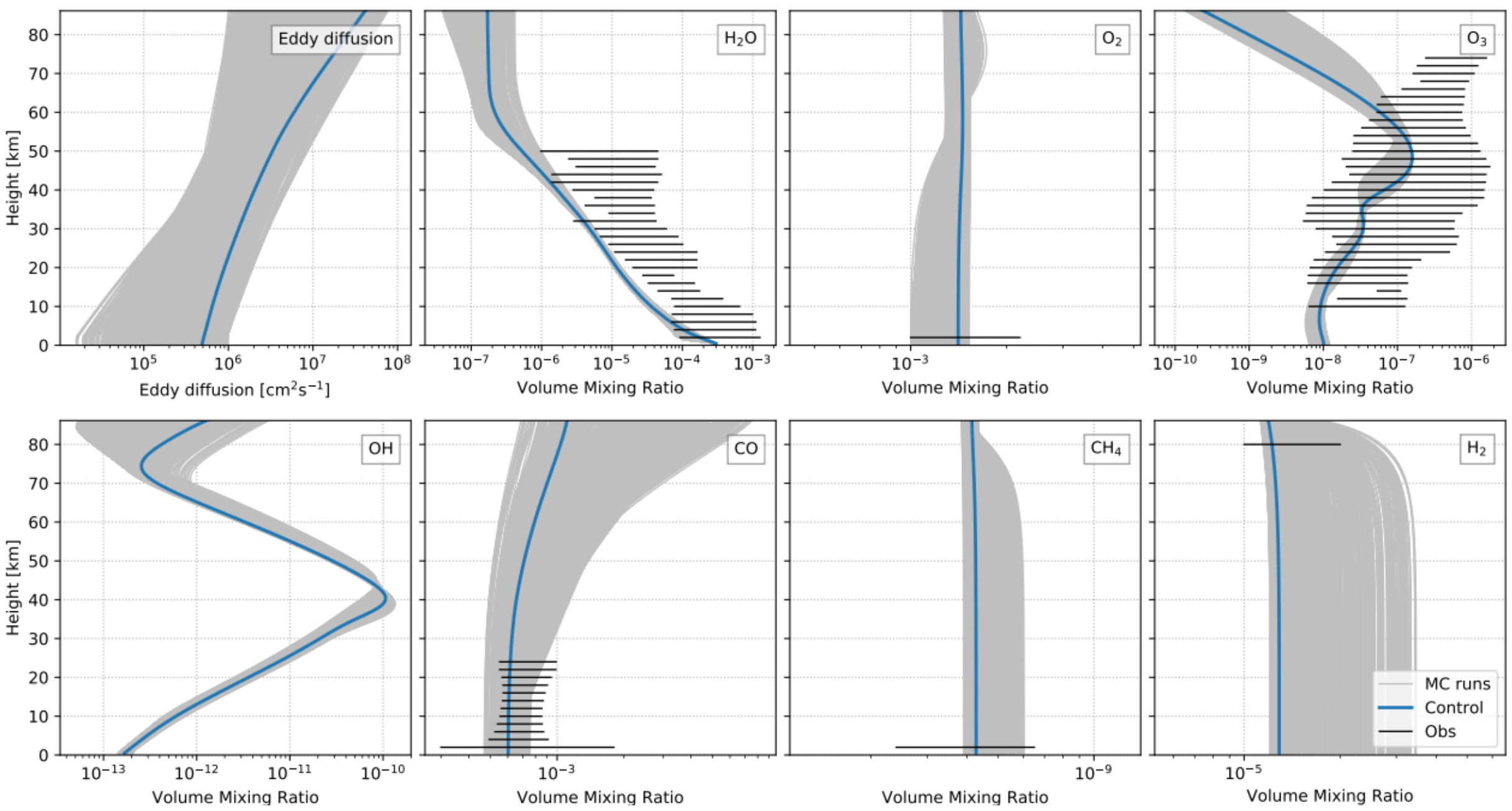


*Figure 4b: as for Figure 4a but for scenario 3.2.2 which varies only the eddy coefficients.*

Figure 4b suggests that the eddy uncertainty accounts for around half or more of the overall uncertainty (compare with Figure 4a) for the longer-lived species e.g. $O_2$, CO, $CH_4$ and $H_2$ since their abundances are determined mainly by both transport and chemical sources. For CO, $CH_4$ and $H_2$ the eddy uncertainty distribution leads to mainly an increase in the abundance whereas for $O_2$ it leads to an increase (decrease) in the upper (lower) layers. The $O_2$ increase in the upper layers was likely related to the $O_3$ increase in the MC runs compared with the control since $O_3$ photolysis is a major source of $O_2$ in this region. The majority of the MC runs for the variation in eddy diffusion coefficient (see Figure 4a, upper left panel) led to a slowing in transport. Since the $O(^3P)$ vmr increased with height then a slowing in mixing would favor a reduction in $O(^3P)$ hence $O_3$. This therefore suggests that a different, chemical response could be responsible for the $O_3$ increase of the MC runs compared with the control (e.g. the increase in $CH_4$ for the MC runs could shield $O_3$ photolysis) although this requires further investigations. In Figure 4b the Monte Carlo variation in the minimum to maximum vmr over the profile of key long-lived species was $O_2$ ($1.0x10^{-3}$-$2.20x10^{-3}$), CO ($3x10^{-4}$-$1.8x10^{-3}$), $CH_4$ ($3.9x10^{-10}$-$7.0x10^{-10}$) and $H_2$ ($1.1x10^{-5}$-$4.0x10^{-5}$). Regarding the observations shown in Figure 4, note that some new data have recently emerged e.g. for CO (Olsen et al., 2021). Note too that the Curiosity rover suggested $N_2$=0.0189 ppb which is lower than the earlier value measured by Viking (=0.027 ppbv; Owen et al., 1977).

Figure 5 shows profiles of the modelled $CH_4$ lifetime, $\tau_{CH4}$ as calculated in BLACKWOLF for the control run (scenario 3.1, solid blue line) and the MC realizations (grey lines). Figure 5a shows the variation due to both eddy and photochemistry uncertainties (scenario 3.2.1) whereas Figure 5b shows the variation due to eddy uncertainty only (scenario 3.2.2).

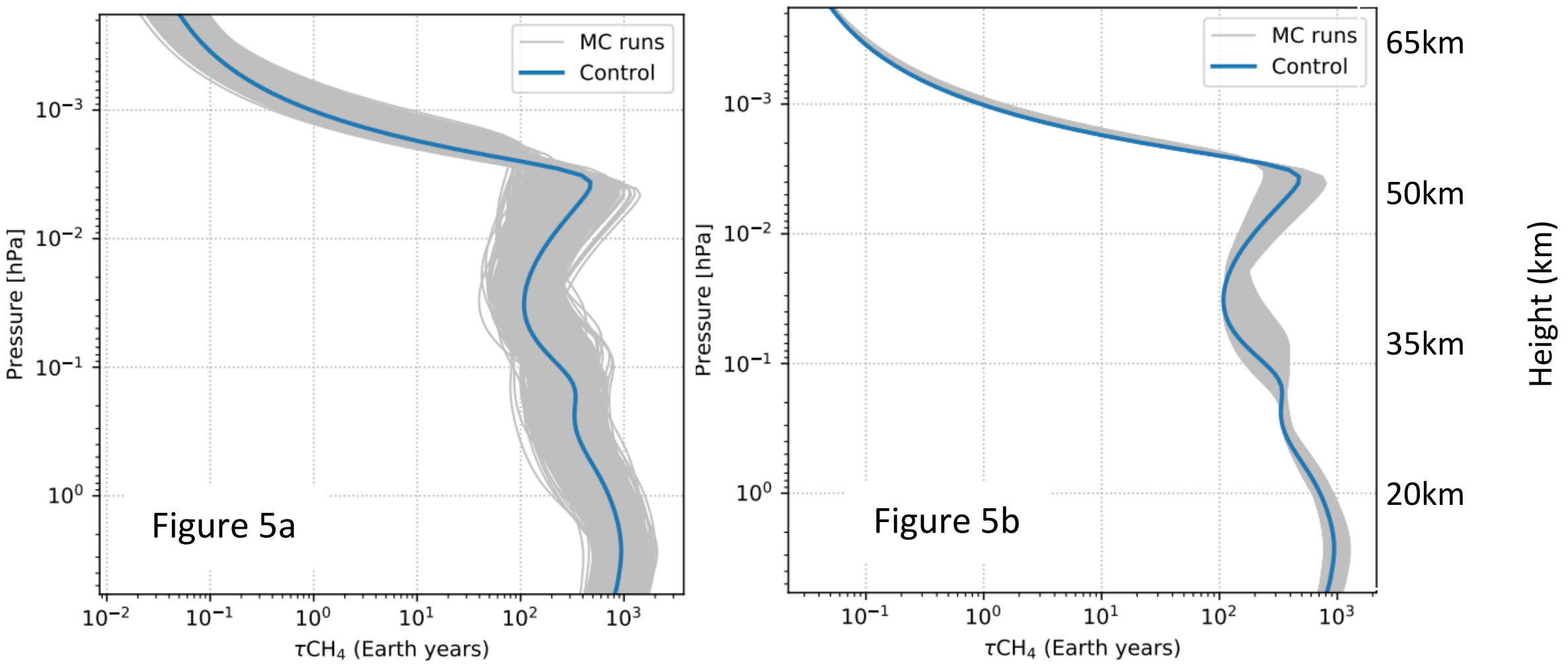


*Figure 5: $CH_4$ atmospheric e-folding lifetime ( $\tau_{CH4}$) in Earth years. The left panel (5a) shows the variation due to both eddy and photochemistry uncertainty (scenario 3.2.1) whereas the right panel (5b) shows the variation due to eddy uncertainty only (scenario 3.2.2).*

Comparing Figures 4a and 4b suggests that most of the uncertainty in the $CH_4$ lifetime arises from the photochemistry especially in the lower layers. The $\tau_{CH4}$ lifetime for the control run (solid blue line) varies from about 900 Earth years near the surface, reducing to about 200 years at $3x10^{-2}$ hPa, then strongly reducing in the upper layers. At the surface in Figure 5b the MC runs led to a variation of (700 to 1000) Earth years.

The behavior in the vertical in Figure 5b reflects the trade-off between the main $CH_4$ sinks (i.e. reactions with OH, $O(^1D)$ and photolysis; refer to Figure 3). On the one hand, these sinks are favored on the upper levels by high UV (leading to $CH_4$ photolysis and the photolytic production of OH and $O(^1D)$) but, on the other hand, favored on the lower levels by increased abundances of their precursors (e.g. $H_2O$ which forms OH). The MC runs (grey lines) in Figure 5a suggest that the MC runs lead to a range in $\tau_{CH4}$ of up to a factor of five, from about 400 to 2000 Earth years in the lower and middle atmosphere, and can decrease by up to a factor of about three from the Mars control value (shown in blue) at the surface.

In general, the mean global atmospheric lifetime ($\overline{\tau}$) of source gases is defined as the atmospheric burden (total density) divided by the integrated loss rate (Brasseur et al., 1999) and can be written as follows:

$$\overline{\tau} = \frac{\int_0^H c(z)\,dz}{\int_0^H R(z)\,dz} \qquad (3)$$

where H=height at TOA, c=species concentration, z=altitude and $R_{CH4}$ is the net rate of atmospheric removal (molecules $cm^{-3}$ $s^{-1}$).

Figure 5 suggests a mean e-folding lifetime $\overline{\tau_{CH4}}$ value of ~400 Earth years (calculated from the harmonic mean weighted by concentration in the BLACKWOLF model, see equation 3). Lefèvre and Forget (2009) (see also Lefèvre (2019)) calculated $\overline{\tau_{CH4}}$=330 Earth years in the Laboratoire de Météorologie Dynamique (LMD) global climate model with photochemistry. Our somewhat higher lifetime value is associated with e.g. our extended chemical network with updated rate coefficients. This impacts the contribution of $CH_4$ removal by e.g. OH (see section 4.1). Such long lifetimes would lead to $CH_4$ spreading evenly over the entire planet. However, observations suggest local spikes of

$CH_4$ (as discussed in the literature overview section 1) which are removed on the order of about 200 Earth days or faster. Lefèvre (2009) suggested that current atmospheric physics and chemistry, assuming $CH_4$ is present, could not account for these observations and would lead to removal about 600 times slower. Figure 6 represents the distribution of the $CH_4$ atmospheric lifetime and a corresponding lognormal fit function for the MC runs from scenario 3.2.1:

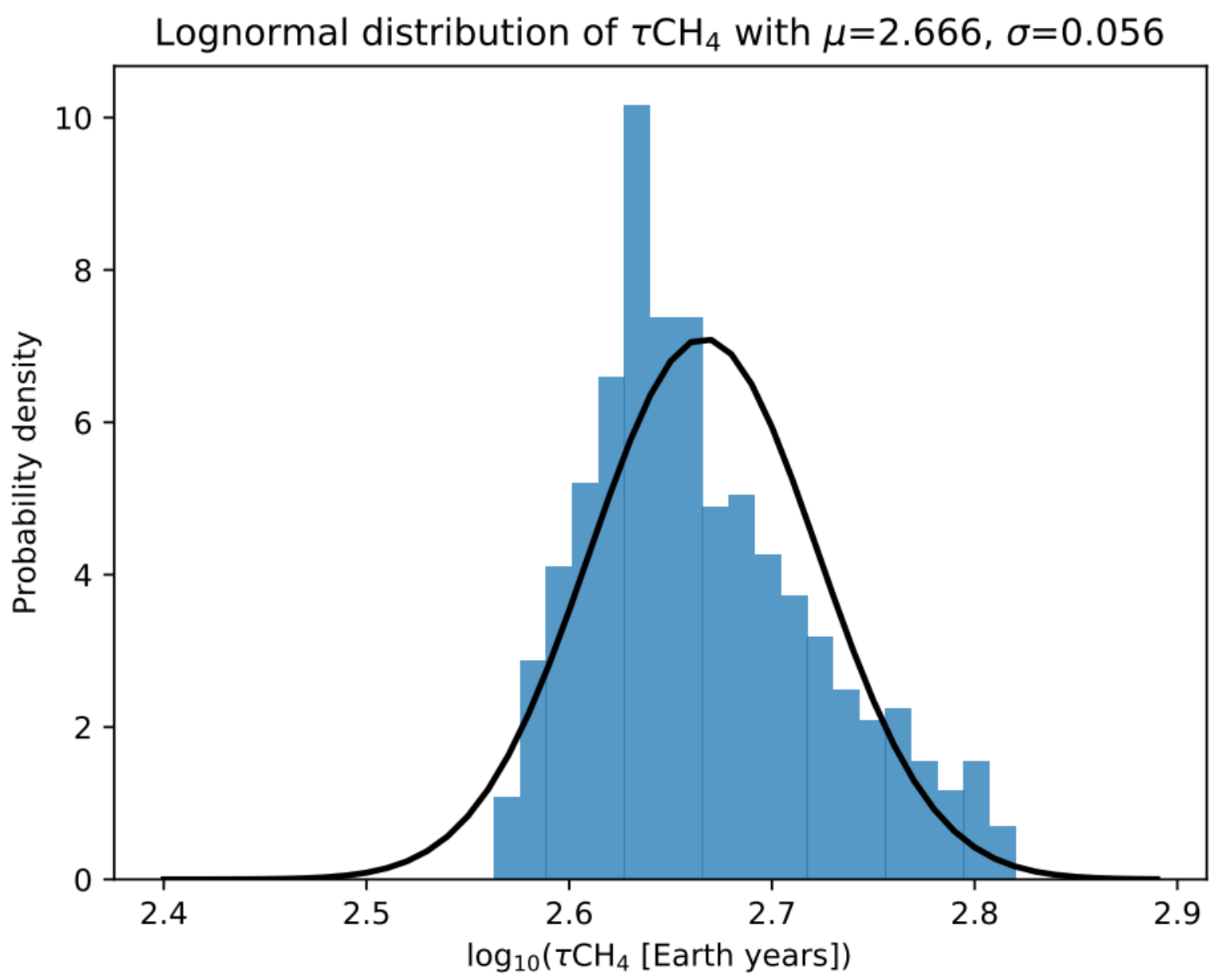


*Figure 6: Probability Density of the $CH_4$ vertically-averaged lifetime (scenario 3.2.1; see Eq. 3). Blue bars denote MC data whereas the solid black line denotes a lognormal fit function.*

Figure 6 shows the (unitless) Probability Density Function (PDF) constructed such that the integral under the curve equals 1. The black fit-line in Figure 6 suggests that $\log_{10}(\tau_{CH4})$ is approximately normally distributed with $\mu$ = 2.666 (=463 Earth years) and $\sigma$=0.056. Reducing the logarithmic lifetime by $2\sigma$ corresponds to a reduction in the lifetime by a factor of $10^{2\sigma}$ ~1.3 which results in a change from 463 to 355 years. For a normal distribution, 95.4% of all datapoints lie within $\log_{10}(\tau_{CH4})$ $\pm 2\sigma$. Consequently, ½·(100-95.4) = 2.3% of all datapoints assume values smaller than $\mu - 2\sigma$. If 'p' denotes probability, this means that p ($\log_{10}[\tau_{CH4}] < \mu - 2\sigma$) = 2.3%, i.e. p ($\tau_{CH4} < 10^{\mu-2\sigma}$) = 2.3%; inserting the values for $\mu$ and $\sigma$ yields: p [$\tau_{CH4}$ < (1/2.65) · 463 years] = 2.3%. This means a reduction in the lifetime by a factor 1.3 has a probability of 2.3%.

### 4.4 Water profile variation (scenario 3.3)

The panels in Figure 7 show species profiles calculated by the BLACKWOLF model with the water profile variation:

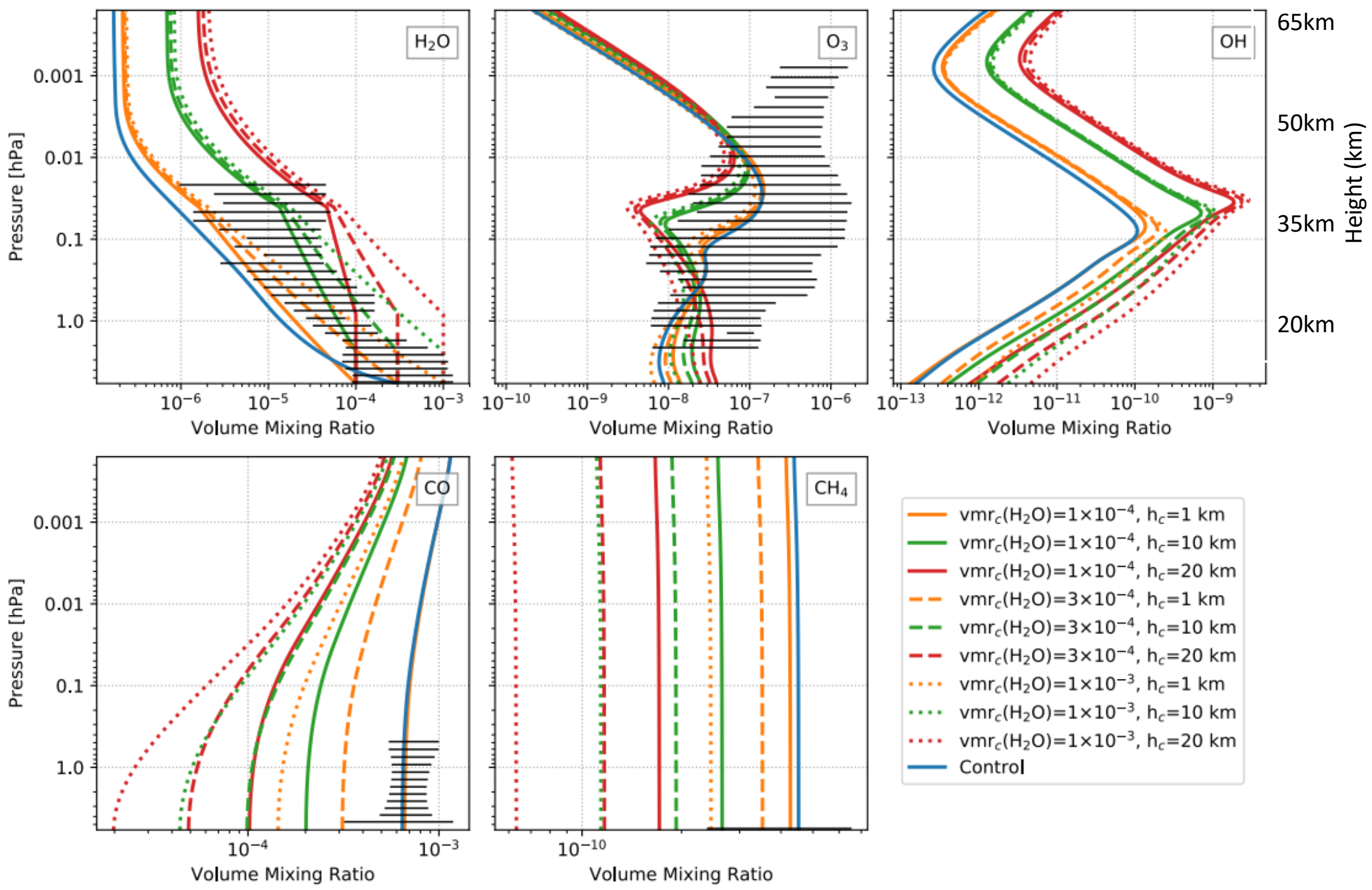


*Figure 7: Species profiles for $H_2O$, $O_3$, OH (upper panel from left to right) and CO, $CH_4$ (lower panel from left to right) (vmr) calculated by the BLACKWOLF model with different input water profiles (scenario 3.3) set to fixed isoprofiles below the level ($h_c$) shown. Observations (horizontal black lines) are as described in Figure 4. The Mars control run (scenario 3.1) is shown as the solid blue line.*

Figure 7 suggests that increased $H_2O$ leads to increased OH (hence HOx=OH+$HO_2$) hence decreased CO and $CH_4$. HOx can catalyze the recombination of CO and O (a precursor for $O_2$ and $O_3$) into $CO_2$ (see e.g. Stock et al., 2012) via:

$$CO + OH \rightarrow CO_2 + H$$
$$H + O_2 + M \rightarrow HO_2 + M$$
$$HO_2 + O(^3P) \rightarrow OH + O_2$$

------------------------------

$$CO + O(^3P) \rightarrow CO_2$$

where ‘M’ refers to any third body required to carry away excess vibrational energy. The CO profile (lower row, right panel) suggests that eight of the runs performed i.e. those with enhanced water abundances, led to OH amounts which suppressed CO below the observed global mean range. Note however that estimating global mean CO on Mars is rather challenging due to e.g. seasonal and annual cycles (e.g. Smith et al., 2018). Results nevertheless give a hint that some of the enhanced water abundances assumed in scenario 3.3 are not favored by the CO observations.

Figure 8 shows the same scenarios as Figure 7 but for $\tau_{CH4}$:

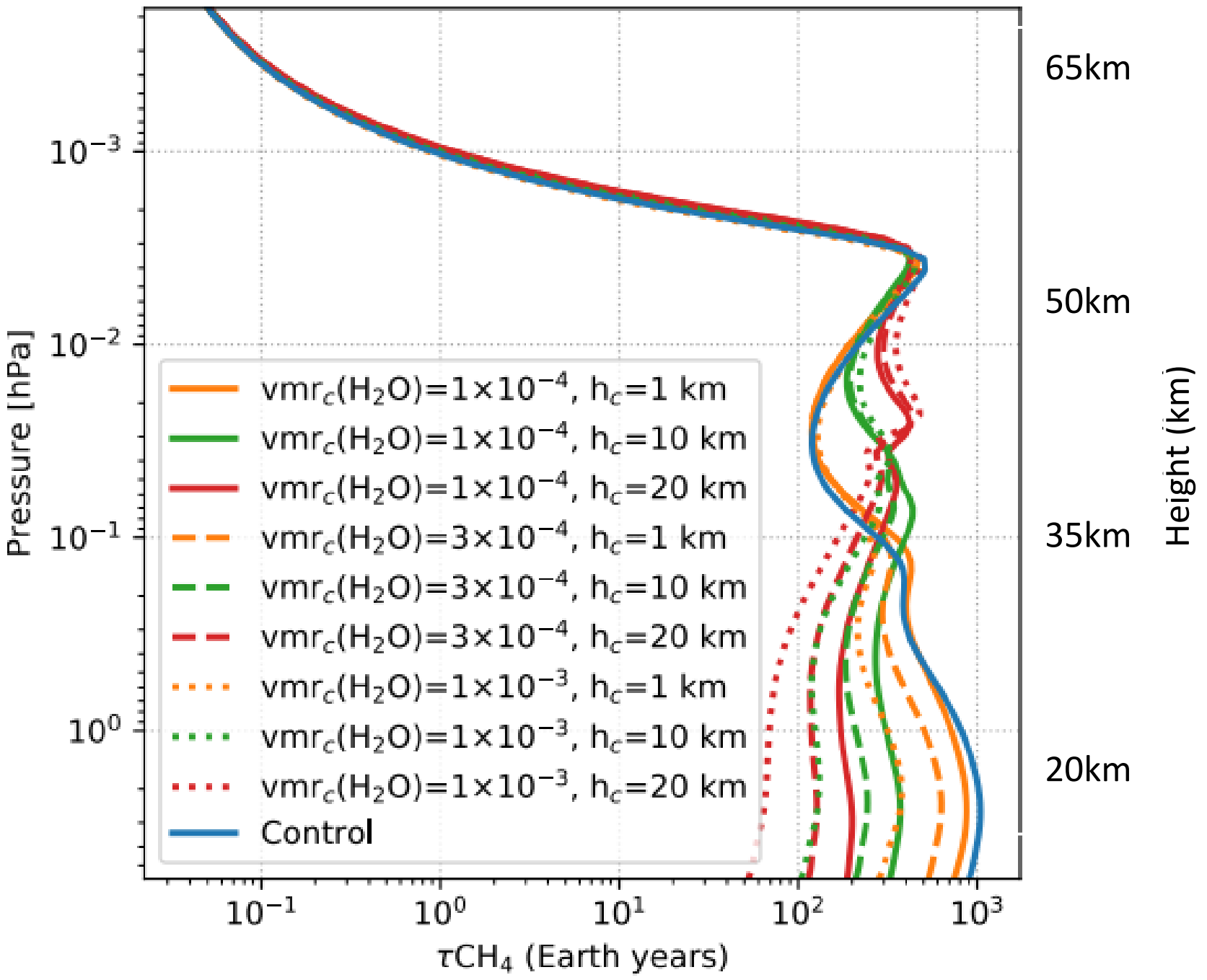


*Figure 8: $\tau_{CH4}$ (Earth years) calculated by the BLACKWOLF model for the water variation runs (scenario 3.3).*

Figure 8 suggests a more substantial effect of $H_2O$ upon $\tau_{CH4}$ in the lower layers. $H_2O$ affects $CH_4$ as follows. In the lower layers, increasing $H_2O$ leads to increases in OH which is an important sink for $CH_4$, whereas in the upper layers the main sinks for $CH_4$ are via reaction with $O(^1D)$ and photolysis (refer to Figure 3), which are less affected by a variation in $H_2O$. Excluding the high water runs (dotted red, dotted green and dashed red lines) in Figure 8 because they are not consistent with the CO observations (see text to Figure 7), Figure 8 suggests that the variation in $\tau_{CH4}$ due to varying $H_2O$ is up to a factor of ~2.1 on comparing the vertical average lifetimes of the solid blue (control) and solid red lines (see also section 5).

**4.5 Chlorine Burden Variation (Scenario 3.4)**

Figure 9 shows species profiles output from BLACKWOLF for the two cases of scenario 3.4 with varying chlorine burdens:

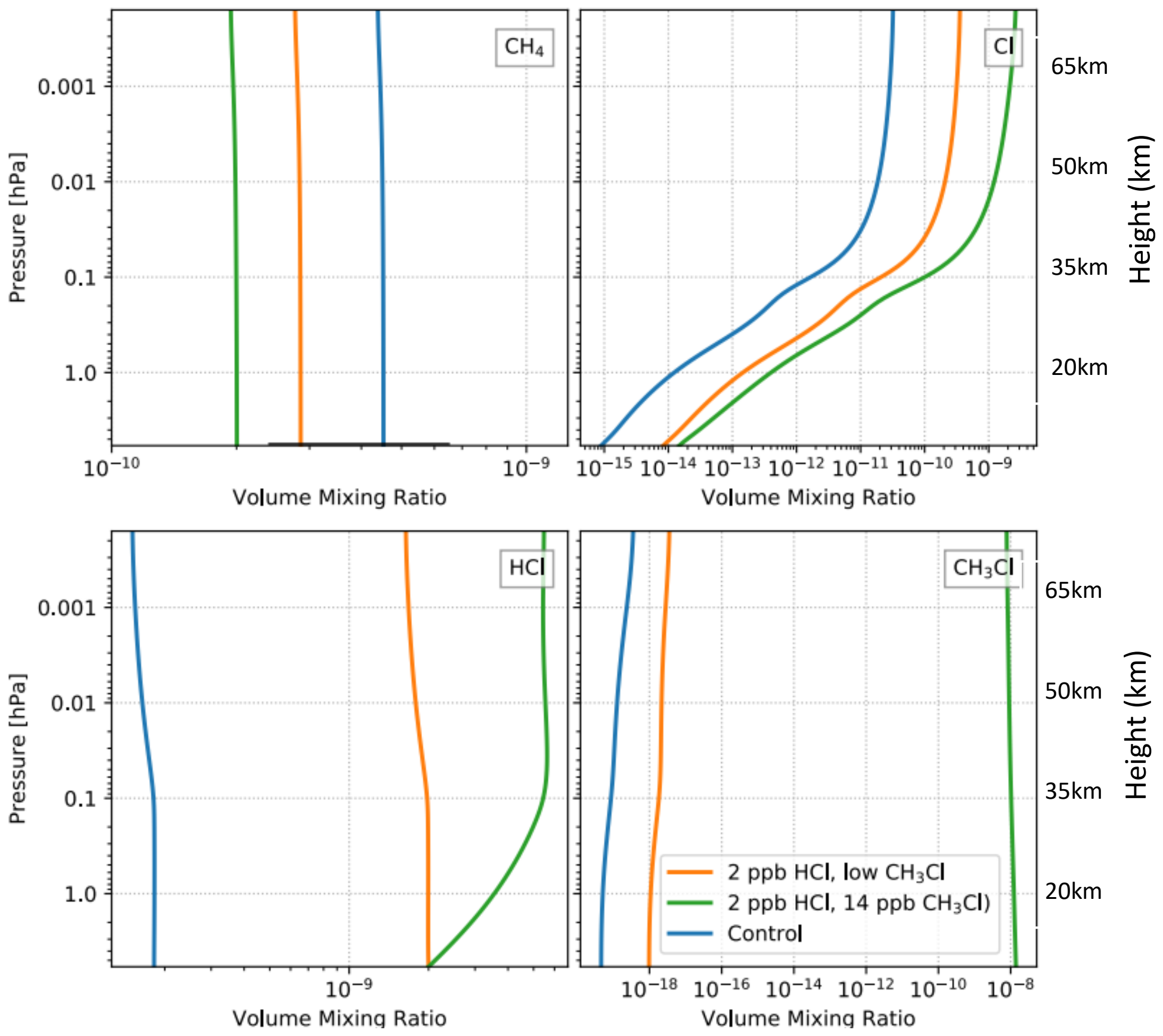


*Figure 9: Species profiles for scenario 3.4 with varying atmospheric chlorine burden calculated in BLACKWOLF. The solid blue line shows the control run (3.1) for comparison (surface HCl fixed to ~0.2ppbv). The orange line shows case 3.4.1 (surface HCl fixed to 2 ppbv and no surface source of $CH_3Cl$) whereas the green solid line shows case 3.4.2 (surface HCl fixed to 2 ppbv and surface $CH_3Cl$ fixed to 14 ppbv based on the upper limit of Villanueva et al., 2013). The four panels show vmr profiles for $CH_4$ (upper left), atomic chlorine Cl (upper right), HCl (lower left) and $CH_3Cl$ (lower right).*

Figure 9 suggests a decrease in $CH_4$ (upper left panel) by around a factor 2.2 in the enhanced chlorine run (solid green line) compared with the control run (solid blue line). Atomic chlorine (upper right panel) which can directly destroy $CH_4$ in the atmosphere increased by up to a factor 100 in the upper layers, HCl (lower left panel) increased by a factor of up to about 30 whereas $CH_3Cl$ (lower right panel) increased strongly by more than ten orders of magnitude compared with the control run. Note that a caveat to Figure 9 is that the high chlorine scenario could be excessive since the assumed $CH_3Cl$ value (=14ppbv) is based on an upper limit observation. The modeled decrease in HCl with decreasing altitude down to the surface could be related to the assumption of fixed HCl abundances at the model surface. Krasnopolsky and Lefèvre (2013) suggest that HCl is formed mainly via: $HO_2$+Cl→HCl+$O_2$. The precursors are photochemically produced (suggesting an increase of HCl with altitude) although the product HCl is photochemically destroyed. Further work is required to investigate the behavior of HCl with altitude.

Figure 10 shows $\tau_{CH4}$ profiles for scenario 3.4 with the varying chlorine burden:

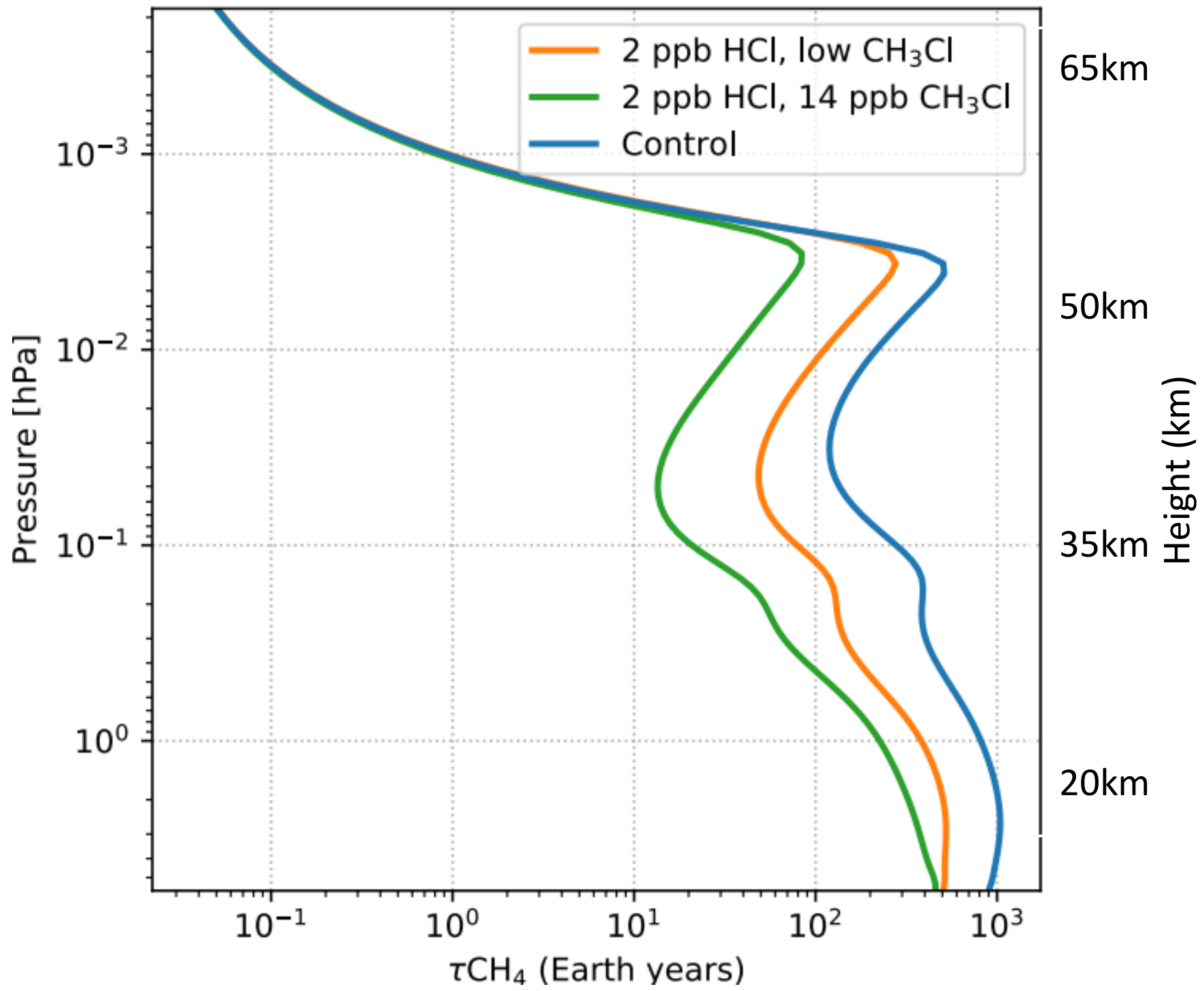


*Figure 10: $\tau_{CH4}$ (Earth years) calculated by BLACKWOLF for the chorine variation runs [cases 3.4.1 (orange line) and 3.4.2 (green line)].*

Figure 10 suggests a reduction in the vertically-averaged lifetime, $\tau_{CH4}$ for the high chlorine burden run (green solid line) compared with the control run (blue solid line) by a factor of ~6.0 with the reduction peaking in the middle atmosphere (see also section 5). Figure 11 is as for Figure 3 showing the removal profiles of $CH_4$ due to the various atmospheric sinks:

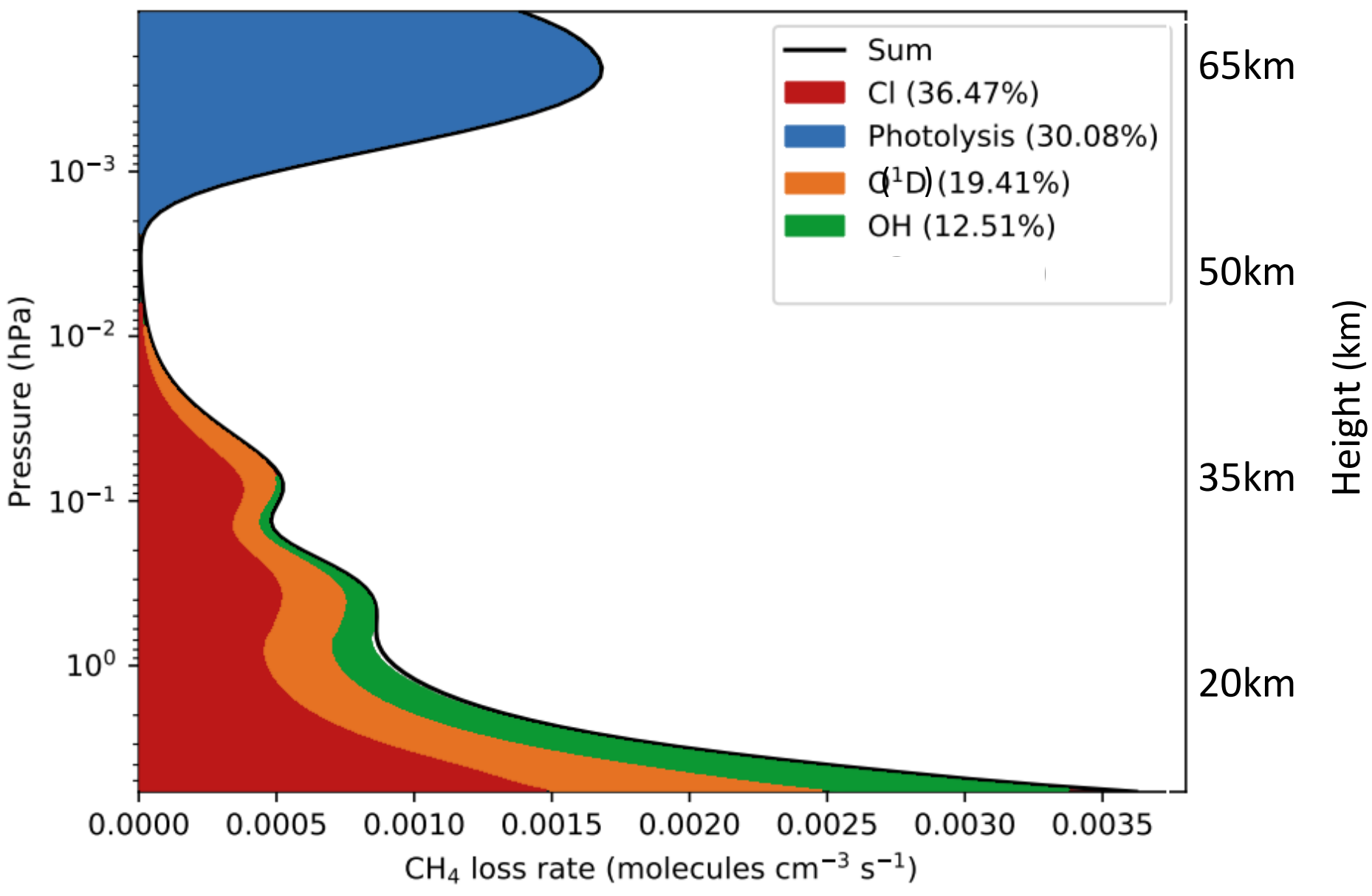


*Figure 11: As for Figure 3 but for the enhanced chorine scenario (3.4.1).*

Figure 11 suggests that the globally-averaged contribution of atomic chlorine (red shading) to the overall atmospheric $CH_4$ removal increases from 5.4% in the control run (see Figure 3) up to 36.5% for the high chlorine run (case 3.4.2).

**4.6 Effect of Low Energy Electrons (Scenario 3.5)**

Figure 12 shows the low energy electron flux profiles derived from the two methods described in Appendix 2b:

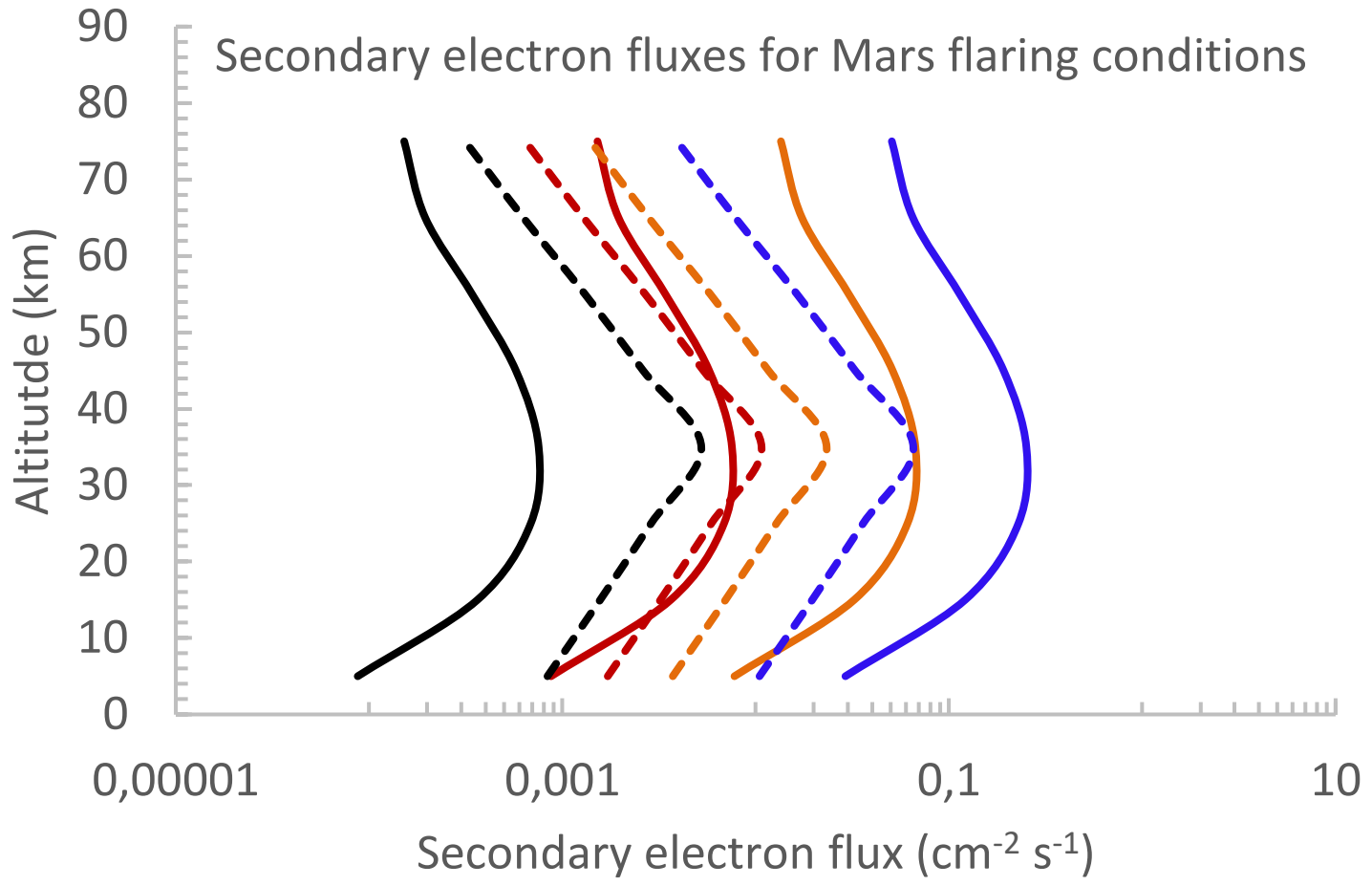


*Figure 12: Low energy secondary electron flux ($cm^{-2}$ $s^{-1}$) profiles derived from method 1 (solid lines) and method 2 (dashed lines) (see Appendix 2b). Black lines denote the energy interval (0.5-5eV); red lines (5-50eV); orange lines (50-500eV) and blue lines (500-5000eV).*

Figure 12 suggests that the derived atmospheric low energy electron flux profiles from the two methods described in Appendix 2b based on 1) the model results by Ehresmann (2012) (solid lines) and 2) multiplying the IPP fluxes by the fraction of low energy electrons ($N_e$) compare reasonably well given the rather simplified scaling assumptions involved. The broadly similar results on comparing both methods in Figure 12 gives confidence that the derived values are reasonable. Both methods are described in more detail in Appendix 2b. In order to test the effect of the secondary electrons in BLACKWOLF however, only the fluxes from method 1 were used to implement electron dissociation (as described in Appendix 2c) since method 1 is based on a more detailed model calculations compared with method 2.

Figure 13 shows the resulting profiles for the Mars control run (solid blue lines) and the scenario with electron dissociation (solid orange lines):

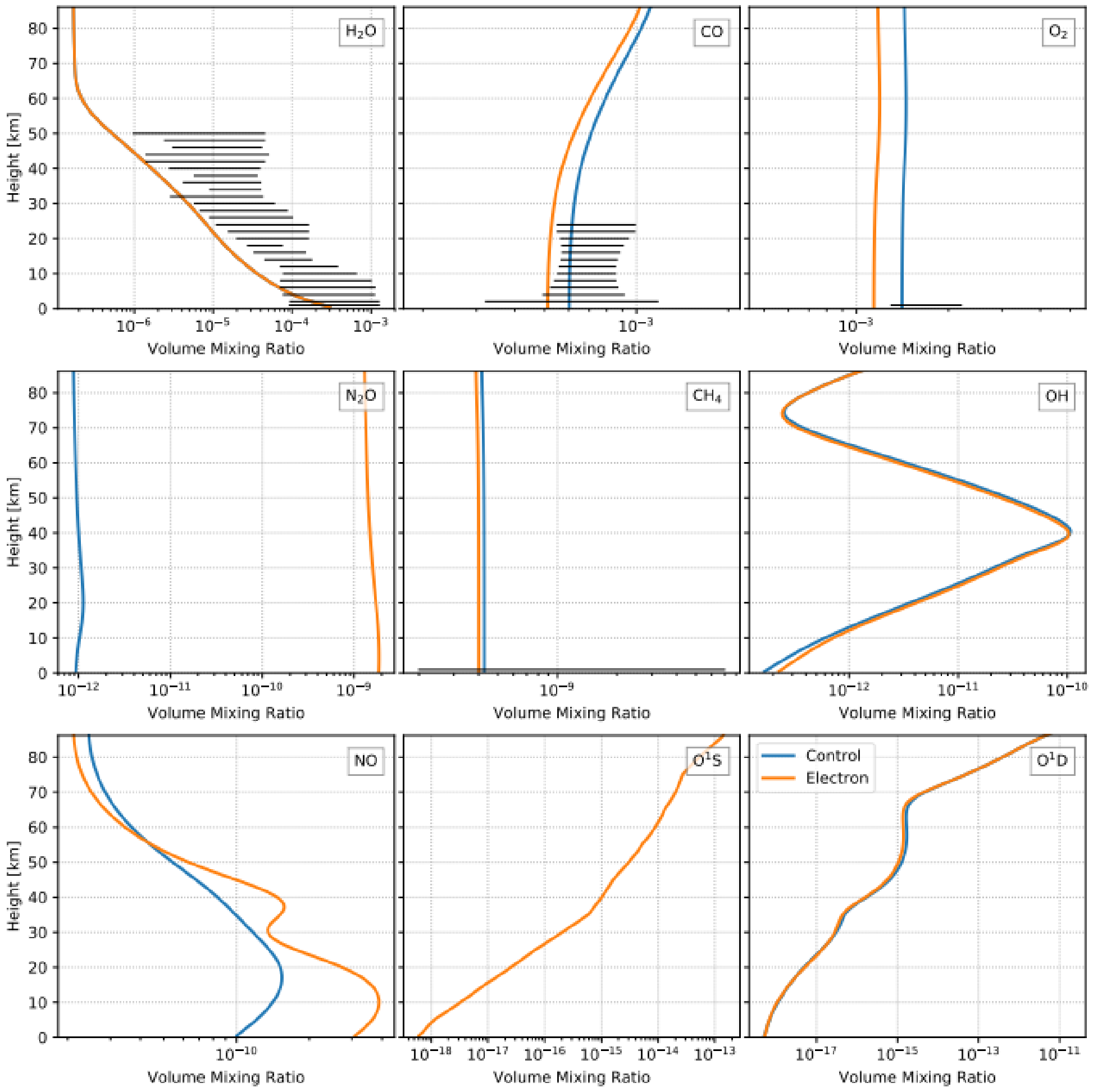


*Figure 13: Species profiles for the Mars control run (scenario 3.1, solid blue lines) and the scenario with electron dissociation (scenario 3.5, solid orange lines). Solid black horizontal lines represent the range of observations (see Wunderlich et al., 2020).*

Figure 13 suggests that the effect of including the electron dissociation rates leads to only a modest reduction of about ten percent in the $CH_4$ vmr. A small amount of $H_2O$ is dissociated electronically which favors a rise in the hydroxyl radical (OH) in the lower layers. $O_2$ is reduced via electronic dissociation by about a factor of two which favors a lowering in electronically excited oxygen ($O(^1D)$) (its photolytic product) in the upper layers and occurs despite the production of $O(^1D)$ via electronic dissociation of $CO_2$. Overall, the reaction $H_2O+O(^1D)$ (an important source of OH) favors OH formation in the lower layers consistent with the $H_2O$ response but disfavors OH formation on the upper layers, consistent with the $O(^1D)$ response. The overall effect upon OH and hence upon $CH_4$ is small. Carbon monoxide (CO) whose main sink is also the reaction with OH also decreases in the lower layers. Interestingly, the potential biosignature nitrous oxide ($N_2O$) increases from about 1 part per trillion (ppt) up to about 2 parts per billion (ppb) associated with abiotic production from increased nitrogen oxides favored by electronic dissociation of $N_2$ into atomic nitrogen. Villanueva et al. (2013) reported an upper limit of <87 ppbv $N_2O$ on Mars at longitude (Ls) = 352° and <65 ppbv at Ls = 83°.

Figure 14 shows the rates of chemical reactions featuring $CO_2$ which are output from BLACKWOLF for scenario 3.5 i.e. including electronic dissociation:

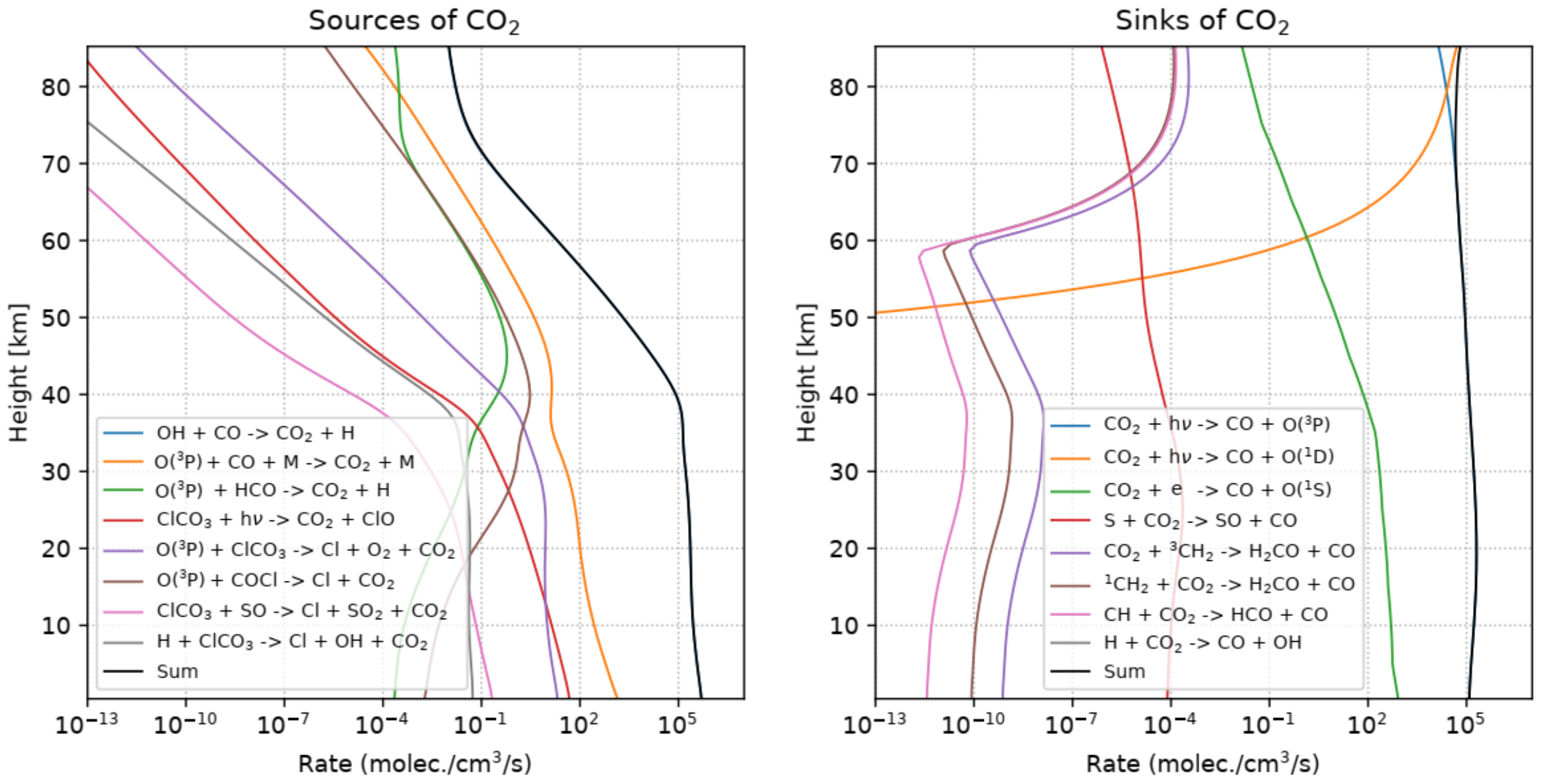


*Figure 14: Rate of $CO_2$ sources (left panel) and sinks (right panel) for scenario 3.5.*

Figure 14 suggests that the $CO_2$ sink due to electronic dissociation (solid purple line, right panel) is about two orders of magnitude slower than the sink due to $CO_2$ photolysis (solid orange line, right panel). Hence, the overall effect of electronic dissociation is rather modest.

### 4.7 Effect of Cosmic Rays

#### 4.7.1 GCRs during solar minimum conditions (Scenario 3.6)

Figure 15 shows the altitude-dependent GCR-induced IPP rates calculated with PLANETOCOSMICS (black line) and AtRIS (purple line) during solar minimum conditions:

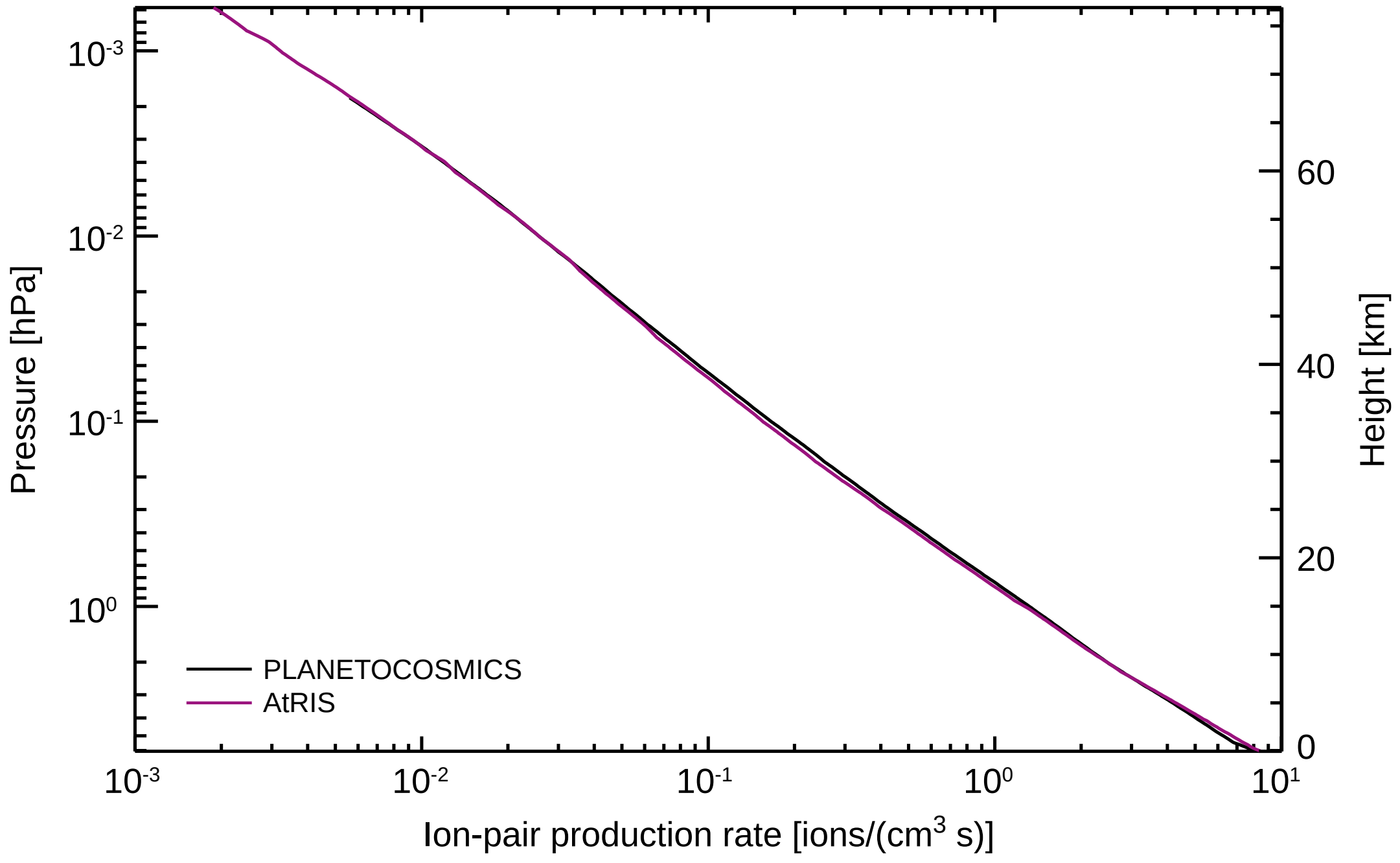


*Figure 15: GCR-induced IPP atmospheric profile rates in the Martian atmosphere for solar minimum conditions (scenario 3.6) calculated with PLANETOCOSMICS (black line) and AtRIS (purple line) based on the proton spectrum by Herbst et al. (2017).*

Figure 15 suggests that on Mars, unlike on Earth, the missing magnetic field generated by the interior and weaker shielding by the thin Martian atmosphere leads to the absence of an ionization maximum in the lower Martian atmosphere. On the contrary our results suggest that the GCR-induced IPP maximum would occur well below the surface. Similar results have been found in studies addressing the Martian surface radiation dose (e.g., Röstel et al., 2020). Further, it shows that the results based on AtRIS are in excellent agreement to those derived with PLANETOCOSMICS.

Figure 16 utilizes GCR-induced IPP profile in the ExoTIC model and shows results for the solar minimum run with GCRs (scenario 3.6), highlighting thermal electron and ion densities as well as the formation and loss of neutral species due to atmospheric ionization:

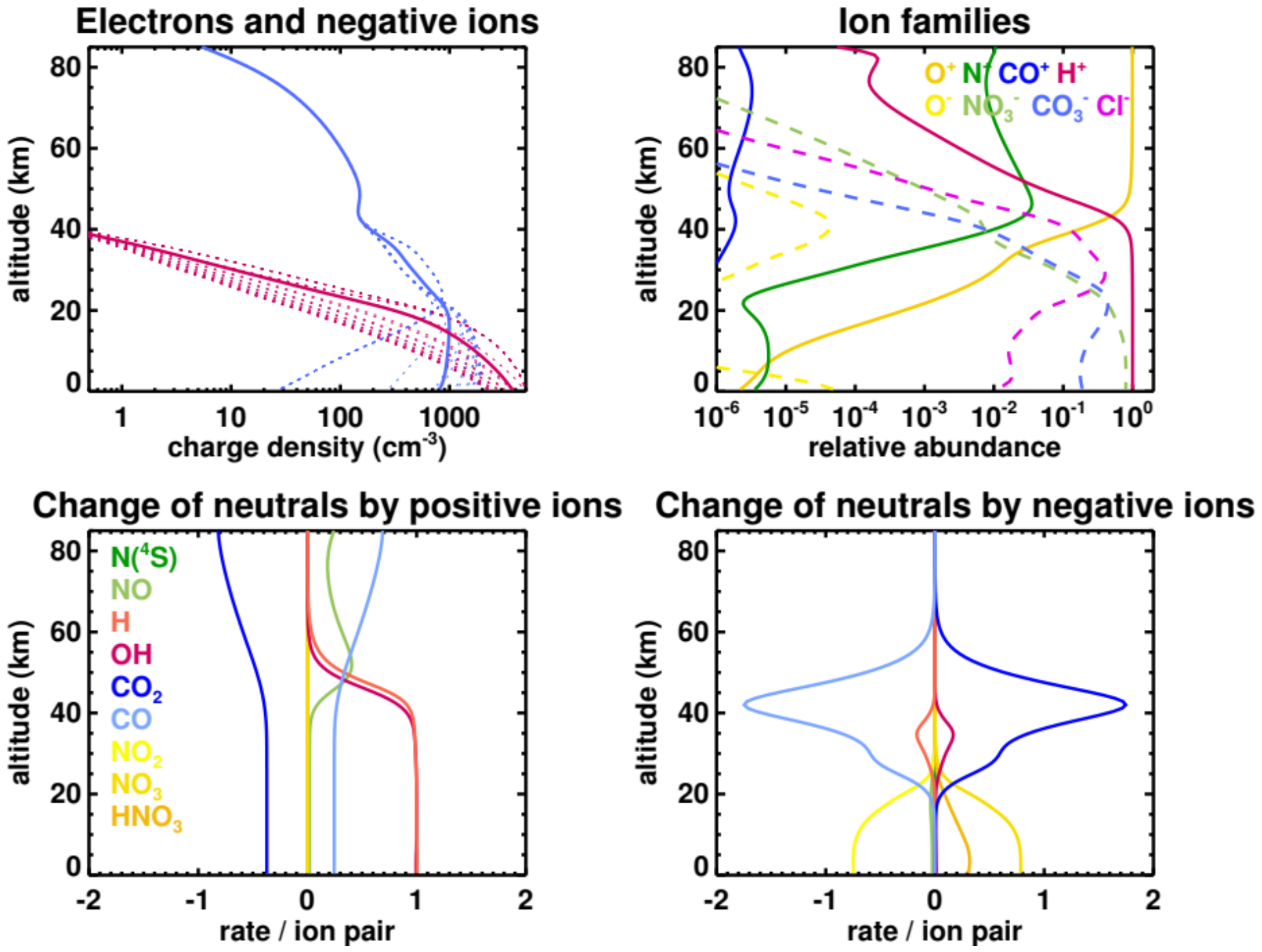


*Figure 16: Thermal electrons from ion-chemical reactions, charged species and formation and loss rates of neutral species due to atmospheric ionization for Mars solar minimum conditions including GCRs (scenario 3.6) as calculated by ExoTIC run for one Mars day (25 Earth hours) with the latitude fixed to 60° and the solar zenith angle calculated depending on the Martian orbital motion (e.g. inclination and rotation) varying over one Martian day with constant ionization conditions. For the GCR case, the background atmosphere in ExoTIC was reset each hour to the value provided by BLACKWOLF each hour before re-calculating the ionization impact. Upper left panel: charge density for thermal electrons (blue) and the sum of all negative ions (red). Solid lines show the mean over one Mars day (25 Earth hours). The 25 thin dotted lines show instantaneous values output at each full hour of the run. Many of the dotted lines overlap however, and are not discernable in the panel. Upper right panel: abundance of ion families (where solid lines show positive ions and dashed lines show negative ions) relative to the total charge abundance. $O^+$: $O^+$, $O_2^+$, $O_4^+$, $O_5^+$. $N^+$: $N^+$, $N_2^+$, $NO^+$, $NO_2^+$, $NO^+N_2$, $NO^+(H_2O)_n$#, $NO_2^+(H_2O)_n$. $H^+$: $H^+$, $H_2O^+$, $H^+(H_2O)_n$. $CO^+$: $CO^+$, $CO_2^+$. $O^-$: $O^-$, $O_2^-$, $O_3^-$, $O_4^-$, $O^-(H_2O)_n$, $O_2^-(H_2O)_n$, $O_3^-(H_2O)_n$. $NO_3^-$: $NO_3^-$, $NO_3^-(H_2O)_n$, $NO_3^-(HCl)$. $CO_3^-$: $CO_3^-$, $CO_3^-(H_2O)_n$. $Cl^-$: $Cl^-$, $Cl_2^-$, $Cl_3^-$, $Cl^-(HCl)$, $Cl^-(H_2O)$, $Cl^-(CO_2)$, $Cl^-(HO_2)$, $ClO^-$. Lower panels show the rate of change of the particular neutral species per ion pair, resulting from (left) reaction chains of positive ions, and (right) reaction chains of negative ions. Values are based on an effective rate i.e. the sum of all production and loss rates and are divided by the ion pair production rate (in $cm^{-3}$ $s^{-1}$), so the quantities shown are unitless. Such values are typically shown in the literature in this form for historical reasons related to e.g. the NOx formation rate (number of NOx per ion pair) for the terrestrial homosphere. #where 'n' refers to the cluster species as follows: $H\cdot(H_2O)$: n=1, ...,7; $NO\cdot(H_2O)$: n=1,2,3; $NO_2\cdot(H_2O)$: n=1,2; coupled clusters: $H\cdot(H_2O)(OH)$; $H\cdot(H_2O)n(CO_2)$, n=1,2; $H\cdot(H_2O)n(N_2)$, n=1, 2; $NO\cdot(CO_2)$; $NO\cdot(N_2)$; $NO(H_2O)\cdot(CO_2)$, n=1,2; $NO\cdot(H_2O)n(N_2)$, n=1,2 and $H\cdot(CH_3CN)m(H_2O)n$, n=0,.., 6; m=1,3.*

Figure 16 (upper left panel) suggests that above 15 km altitude, negative charge is dominated by electrons (blue lines), by negative ions below 15 km (red lines), although absolute numbers and

the partitioning between electrons and negative ions vary significantly below 20 km with solar zenith angle due to photoelectron reactions. Figure 16 (upper right panel) suggests that the main negative ions are (from the surface upwards) $NO_3^-$, $CO_3^-$ and $Cl^-$ cluster ions. The main positive ions are $H^+$ water cluster ions below 45 km and $O^+$ containing ions above 45 km. Figure 16 (lower left panel) suggests OH (red line) and CO (light blue line) are influenced mostly by positive ions in the lower and upper atmosphere respectively. Figure 16 (lower right panel) suggests that negative ions lead to $CO_2$ production (royal blue line) and CO destruction (light blue line) peaking at around 40 km.

The ion distributions in Figure 16 are comparable to Molina-Cuberos et al. (2001) insofar as the main positive ions are water cluster ions (summarized as $H^+$ in Figure 16), and $CO_3^-$ containing ions (summarized as $CO_3^-$, see Figure caption) contribute significantly to the negative charge. The impact of ionization and ion chemistry on the neutral composition differs from Earth in several important ways: While the formation of HOx due to positive water cluster ion reactions is comparable to Earth, the NOx formation, which dominates the ion impact on Earth over a broad altitude region from the lower thermosphere to the mid-stratosphere is nearly negligible on Mars due to the low amounts of $N_2$. In the lower atmosphere below 20 km, uptake of $NO_2$ in negative $NO_3^-$ cluster ions appears to be an important process leading to the formation of $HNO_3$, possibly due to a similar reaction pathway as in the terrestrial stratosphere and mesosphere above 30 km altitude. The conversion of $CO_2$ to CO due to dissociative ionization of $CO_2$ is an important process on Mars but not on Earth; it is however counteracted by uptake of CO into negative ions leading to reformation of $CO_2$ in 20-60 km altitude.

Tests were performed implementing the GCR IPP profile from Figure 15 into the BLACKWOLF chemistry model and including the chemical production efficiency (CPE) profiles output from the Exotic neutral-ion model for the species: NO, $NO_2$, $NO_3$, $HNO_2$, $HNO_3$, $N_2$, $N_2O$, $N_2O_5$, H, OH, $HO_2$, $H_2$, $H_2O$, $O(^3P)$, $O(^1D)$, $O_2$, $O_3$, Cl, ClO, HCl, HOCl, $ClNO_2$, $ClNO_3$, $Cl_2$, CO, $CO_2$, $CH_3$, $HCO_3$, $CH_3CN$ and $H_2SO_4$ in the model setup as described in Herbst et al. (2019)[a]. Results however suggested only a small change in composition of up to a few percent in the species profiles in the BLACKWOLF climate-chemistry profiles (not shown).

### 4.7.2 Strong Ground Level Enhancement (GLE) (GLE05, Carrington Event, and AD774/775 | Scenario 3.7)

It is known that GLE events have a more pronounced impact on the thin Martian atmosphere than, on, e.g., the thick $CO_2$ dominated Venusian atmosphere (e.g. Herbst et al., 2019[b]). This is particularly so in the case of the strongest GLEs such as the GLE05, the Carrington event, and the AD774/775 event. The direct impact on the altitude-dependent IPP rates of such an event penetrating the Martian atmosphere during solar minimum conditions is shown in Figure 17:

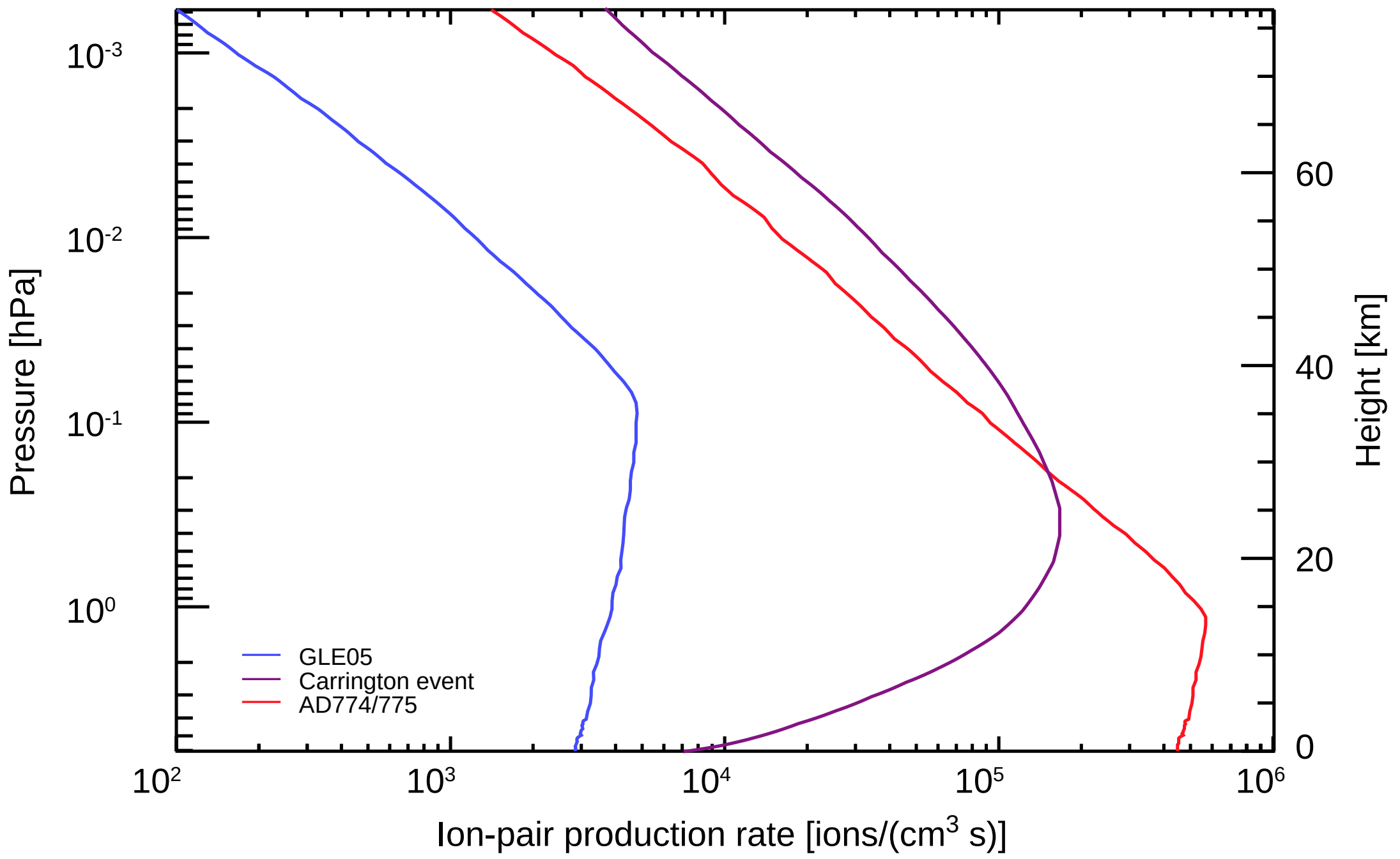


*Figure 17: IPP (ions cm$^{-3}$ s$^{-1}$) profiles in the Martian atmosphere induced by GLE05 (blue line), a Carrington-like event (purple line), and an AD//4/775-like event (red line) modeled with AtRIS (Banjac, Herbst, Heber, 2019). See scenario 3.6 for further information.*

Figure 17 suggests that the GLE-induced IPP rates at the Martian surface are more than two (GLE05) and four (AD774/775) orders of magnitude higher than during solar quiet times dominated by GCRs only (see Figure 15). It can also be seen, that GLE05 would have peaked around 0.1 hPa showing a maximum IPP rate of about 4 x $10^3$ ions/(cm$^3$ s), while the Carrington event and an AD774/775-like event would have had a much stronger impact on the Martian atmospheric ionization. The latter for example, would have impacted the ionization down to about 1hPa with a maximum IPP rate of 5.6 x $10^5$ ions/(cm$^3$ s). This maximum is induced by the much higher amount of low-energy solar energetic particles impinging on and being absorbed by the Martian atmosphere.

Figure 18 shows the corresponding NOx and HOx responses calculated by the ExoTIC model:

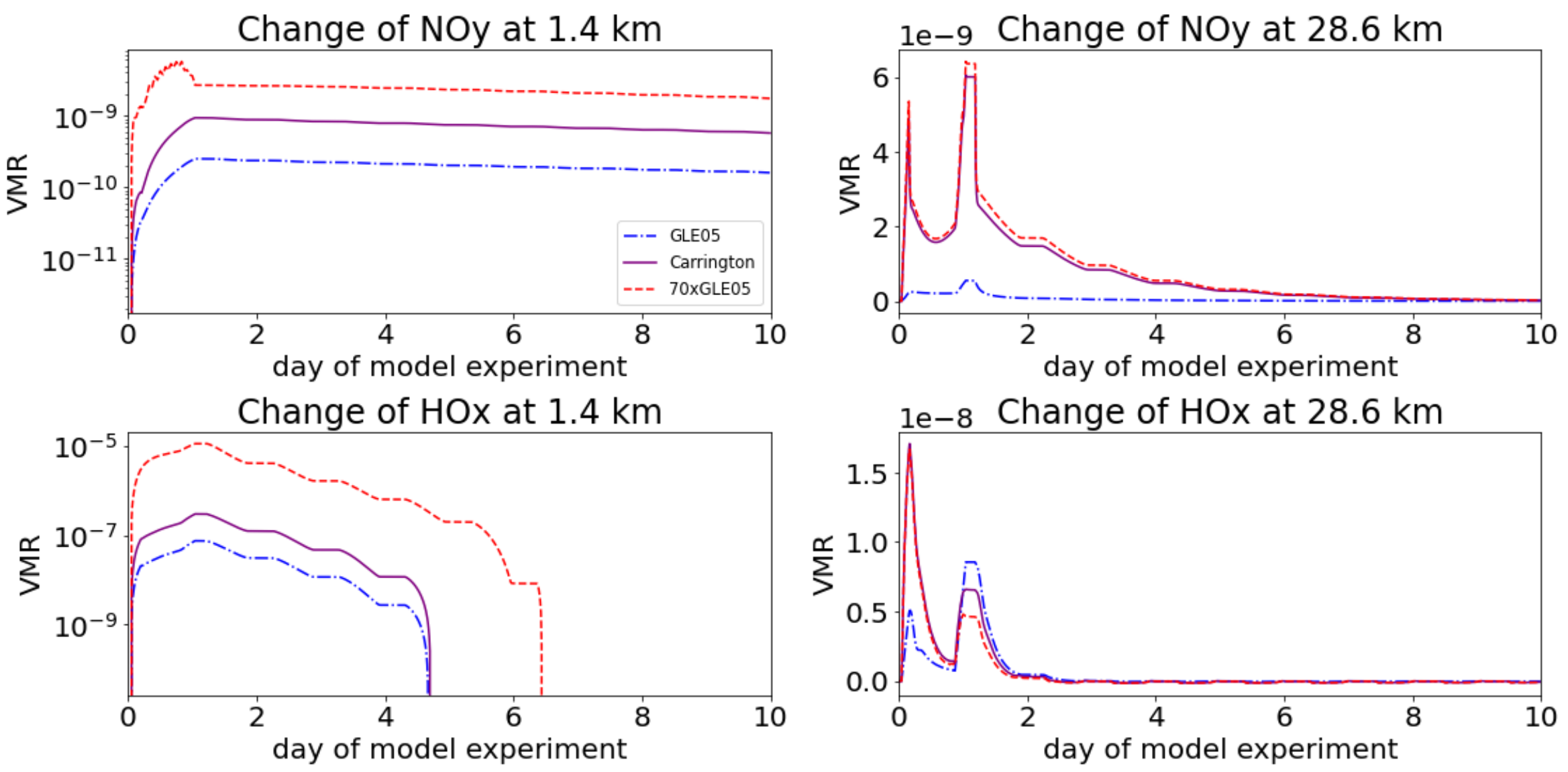


*Figure 18: Time series for the Mars SEP flaring events (scenario 3.7). The y-axis shows the difference (vmr) of two runs with the ExoTIC model for (with ionization – without ionization) for NOy (=N+NO+$NO_2$+$NO_3$+2$N_2O_5$+$HNO_3$+$HNO_4$+Cl$NO_3$) (upper panels) and HOx (=H+OH+$HO_2$+ 2 $H_2O_2$) (lower panels) at 1.4km (left) and 28.6 km (right). The x-axis shows time after beginning of the model experiment in Mars days. Ionization due to the GLE (Carrington event) was imposed for 24 hours starting on the second hour of day one of the model run at a position of 60° N and 0°E. For the GLE case (unlike the GCR case, see above), the background atmosphere in ExoTIC was not reset each hour, in order to investigate the change in the neutral species over time.*

Figure 18 suggests a modest decrease in NOy (up to several ppbv for the AD774/775-like event) and a rather strong (ppbv to several ppmv depending on scenario) increase in HOx near at 1.4 km (left panels) and a moderate increase (up to several ppbv) in both NO**y** and HOx in the altitude of the largest ionization rates (28.5 km, right panels). It is interesting to note that the AD774/775-like scenario and the "Carrington" scenario are very similar at 28.6 km, but differ substantially at the surface due to the harder spectrum of the AD774/775-like scenario. Further investigation suggested that the near surface HOx increase was associated mainly with an enhancement in the hydroperoxy ($HO_2$) radical and its direct conversion into hydrogen peroxide ($H_2O_2$). The chemical network did not favor a strong increase in OH. This suggests that OH as a source of strong methane loss due to GLEs is not likely. Compared to Earth, Mars features a strong CO abundance which via: CO + OH → $CO_2$ + H is an efficient sink for OH and also a source for H (which can quickly form $HO_2$ via: H + $O_2$ + M → $HO_2$ + M). This effect favors $HO_2$ being the dominant HOx species on Mars in the lower atmosphere (see discussion in Lefèvre and Krasnopolsky, 2017). Due to the low amount of $N_2$ available in the Mars atmosphere compared to Earth, the increase in NOy is comparatively small compared to similar events on Earth and is partly overcompensated for at the surface by a solar-zenith angle dependent uptake of $NO_2$ into negative ions consistent with the formation/loss rates of neutrals shown for the GCR case in Figure 16. There was also an increase in ClOx (Cl+ClO+HOCl+OClO+Cl$NO_3$) due to the flaring events although the effect was very small (up to several ppt at 1.4km and up to a few tenths of ppt at 28.6 km) (not shown). It should be noted that the changes to atmospheric composition are not long-lasting; after nine days, the composition of the short-lived HOx, NOx and ClOx compounds has recovered to background values with the exception of NOy at the surface, which is longer-lived.

## 5. Conclusions

If methane exists on Mars then current knowledge of atmospheric physics and chemistry cannot explain its atmospheric responses. Our uncertainty analysis suggests a potential lowering in the simulated methane lifetime due to the following factors:

(1) Factor 1.3 due to photochemical and Eddy coefficient uncertainty (see text to Figure 6)
(2) Factor 2.1 due to changes in the water column (see text to Figure 8)
(3) Factor of 6.0 on adopting the recently suggested high atmospheric chlorine abundances (see text to Figure 10)

These three effects can therefore together account for a lowering in the $CH_4$ lifetime by a factor, $F_{tau}$ where $F_{tau}$ = 1.3*2.1*6.0 = 16.38

Note that the above estimate represents the upper limit of the statistical analysis performed. Furthermore, some of the abundances assumed in our global, annual mean model study (e.g. chlorine) were based on local measurements and could therefore be overestimated in this work. Note too that the upper abundance assumed for the global mean $H_2O$ profile in this work could be an over-estimate (see Villanueva et al., 2021). The above result suggests that, if methane exists on Mars then there remains a significant over-estimate in the modelled $CH_4$ lifetime due to e.g. poorly constrained atmospheric abundances or/and unknown atmospheric and surface sinks.

The abundance of methane on Mars is still debated and could be smaller than the Curiosity levels assumed in the scenarios of the current work. A further caveat is that there are additional uncertainties in our atmospheric column model due to e.g. the climate parameterizations (line-fitting procedure, collision-induced absorption parameterization etc.) and in the chemistry (escape parameters etc.) which are difficult to quantify and beyond the scope of our work. Furthermore:

- Our analysis suggested that methane is mostly removed via photolysis at pressures less than about 0.01 hPa and mainly via atomic chlorine, hydroxyl and excited atomic oxygen at greater pressures.

- Dissociation by low energy electrons does not strongly affect methane but could impact the potential biosignature nitrous oxide although more electron cross section data are needed.

- The effect of galactic cosmic rays upon atmospheric composition is small, up to a few percent.

- The effect of strong ground level events in HEPs leads to a change (at 1.4km height) in NOy of up to a few ppbv and up to 10 ppmv in HOx. These changes typically last for up to a few Earth days.

**Acknowledgements**

JLG and KH thank ISSI Team 464 for fruitful discussion. KH thanks Dr. S. Banjac and Prof. Dr. B. Heber from the Christian-Albrechts-Universität zu Kiel for providing and maintaining AtRIS. MS and FW acknowledge support from DFG projects RA 714/9-1 and RA-714/7-1 respectively. JLG, MS and KH acknowledge the support of the DFG priority program SPP 1992 "Exploring the Diversity of Extrasolar Planets (GO 2610/2-1 and HE 8392/1-1)".

## Appendix 1: Uncertainty Ranges

### Appendix 1a: Bimolecular Reaction Coefficients

Table A1a shows the uncertainty range for the bimolecular reaction coefficients used in the Monte Carlo simulation (MC-1) calculated for 212K for illustration:

| Reaction | Rate Constant (molecule$^{-1}$ cm$^3$ s) | Uncertainty Range* |
|---|---|---|
| $H+HO_2 \rightarrow OH+OH$ | $7.2 \times 10^{-11}$ | 1.37 |
| $H+HO_2 \rightarrow O(^3P)+H_2O$ | $1.6 \times 10^{-12}$ | 1.72 |
| $H+HO_2 \rightarrow H_2+O_2$ | $6.9 \times 10^{-12}$ | 1.60 |
| $H+O_3 \rightarrow OH+O_2$ | $6.9 \times 10^{-10}$ exp(-470/T) | 1.23 |
| $HO_2+HO_2 \rightarrow H_2O_2+O_2$ | $3.5 \times 10^{-13}$ exp(430/T) | 1.32 |
| $HO_2+NO \rightarrow OH+NO_2$ | $3.5 \times 10^{-12}$ exp(250/T) | 1.10 |
| $HO_2+O_3 \rightarrow OH+O_2+O_2$ | $1.0 \times 10^{-14}$ exp(-490/T) | 1.28 |
| $O(^3P) +H_2O_2 \rightarrow OH+HO_2$ | $1.4 \times 10^{-12}$ exp(-2000/T) | 1.37 |
| $O(^3P)+HO_2 \rightarrow OH+O_2$ | $3.0 \times 10^{-11}$ exp(200/T) | 1.12 |
| $O(^3P)+NO_2 \rightarrow NO+O_2$ | $5.0 \times 10^{-12}$ exp(210/T) | 1.14 |
| $O(^3P)+OH \rightarrow O_2+H$ | $2.2 \times 10^{-11}$ exp(120/T) | 1.23 |
| $O(^1D)+CH_4 \rightarrow CH_3+OH$ | $1.3 \times 10^{-10}$ | 1.19 |
| $O(^1D)+CO_2 \rightarrow CO_2+O(^3P)$ | $7.5 \times 10^{-11}$ exp(115/T) | 1.18 |
| $O(^1D)+H_2 \rightarrow H+OH$ | $1.2 \times 10^{-10}$ | 1.23 |
| $O(^1D)+H_2O \rightarrow OH+OH$ | $1.63 \times 10^{-10}$ exp(60/T) | 1.11 |
| $OH+CH_4 \rightarrow CH_3+H_2O$ | $2.45 \times 10^{-12}$ exp(-1775/T) | 1.26 |
| $OH+CO \rightarrow CO_2+H$ | $5.4 \times 10^{-14}$ *(T/298)**1.5*exp(250/T) | 1.09 |
| $OH+H_2 \rightarrow H_2O+H$ | $2.8 \times 10^{-12}$ exp(-1800/T) | 1.20 |
| $OH+H_2O_2 \rightarrow H_2O+HO_2$ | $2.9 \times 10^{-12}$ exp(-160/T) | 1.22 |
| $OH+HO_2 \rightarrow H_2O+O_2$ | $4.8 \times 10^{-11}$ exp(250/T) | 1.23 |
| $OH+O_3 \rightarrow HO_2+O_2$ | $1.7 \times 10^{-12}$ exp(-940/T) | 1.23 |

Table A1a: Bimolecular Rate Coefficient Uncertainties for first Monte Carlo Simulation (MC-1).
*Dividing and multiplying the rate coefficient by the uncertainty range
Respectively gives the $2\sigma$ (95%) error range.

### Appendix 1b: Termolecular Reaction Coefficients

Table A1b is as for A1a but for the termolecular Rate Coefficient Uncertainties.

| Reaction | Uncertainty limit* |
|---|---|
| $H+H+M \rightarrow H_2+M$ | 4.5 |
| $H+O(^3P)+M \rightarrow OH+M$ | 4.9 |
| $H+O_2+M \rightarrow HO_2+M$ | 6 |
| $H+OH+M \rightarrow H_2O+M$ | 4.5 |
| $OH+OH+M \rightarrow H_2O_2+M$ | 2 |

Table A1b: as for A1a but for the termolecular Rate Coefficient Uncertainties. 'M' refers to any third body required to carry away excess vibrational energy. Rate coefficients are rather complex functions of T, p and composition and are listed in JPL (2020).

[*]Based on the upper limit in Nagy et al. (2015) (their Table 3) and denoting the maximum uncertainty in the rate coefficient associated with changing from an $N_2$ to a $CO_2$ bath gas.

## Appendix 2: Gas-phase neutral dissociation by low energy secondary electrons

### Appendix 2a: Electron Cross sections

Table A2a presents cross sections for neutral electron dissociation (shown in black) which were implemented into the BLACKWOLF chemical model.

| Species | $\sigma_n$ (0.5-5.0) eV<br>$\sigma_t$ [0.5-5.0] eV | $\sigma_n$ (5.0-50) eV<br>$\sigma_t$ [5.0-50] eV | $\sigma_n$ (50-500) eV<br>$\sigma_t$ [50-500] eV | $\sigma_n$ (500-5000) eV<br>$\sigma_t$ [500-5000] eV | Reference |
|---|---|---|---|---|---|
| $N_2$ | | $5.9 \times 10^{-17}$<br>[$5.7 \times 10^{-17}$] | $5.3 \times 10^{-17}$<br>[$7.3 \times 10^{-17}$] | | Cosby (1993)[*]<br>Märk (1975)[**] |
| $O_2$ | $4.0 \times 10^{-18}$ | $8.6 \times 10^{-18}$<br>[$3.5 \times 10^{-17}$] | $1.1 \times 10^{-18}$<br>[$9.0 \times 20^{-17}$] | | Maeda and Aikin (1968)[***]<br>Märk (1975)[****] |
| $O_3$ | See legend[#]<br>[$1.5 \times 10^{-17}$] | See legend[#]<br>[$3.0 \times 10^{-19}$] | See legend[#] | See legend[#] | See legend[#]<br>Rangwala et al. (1999)[##] |
| $N_2O$ | See legend[###] | See legend[###]<br>[$1.3 \times 10^{-16}$] | See legend[###]<br>[$2.9 \times 10^{-16}$] | See legend[###]<br>[$7.6 \times 10^{-17}$] | See legend[###]<br>Iga et al. (1996)[####] |
| $CH_3Cl$ | See legend[$]<br>[$1.1 \times 10^{-23}$] | See legend[$] | See legend[$] | See legend[$] | See legend[$]<br>Pearl et al. (1995)[$$] |
| $PH_3$ | | | [$3.1 \times 10^{-16}$] | | Kumar (2014)[$$$] |
| $NH_3$ | | | [$8.8 \times 10^{-16}$] | | Zecca et al. (1992)[$$$$] |
| $CH_4$ | | $1.4 \times 10^{-16}$ | $1.6 \times 10^{-16}$<br>[$3.0 \times 10^{-16}$] | | Shirai et al. (2002)[§]<br>Orient and Srivastava (1987)[§§] |
| $H_2$ | | $1.0 \times 10^{-18}$ | $2.5 \times 10^{-19}$ | | Padovani et al. (2018)[§§§] |
| $H_2O$ | | $1.0 \times 10^{-16}$<br>$4.7 \times 10^{-19}$ | $1.5 \times 10^{-16}$<br>$1.2 \times 10^{-18}$ | | Itikawa and Mason (2005)[§§§§]<br>Itikawa and Mason (2005)[&]<br>See text[&&] |
| $CO_2$ | | $3.2 \times 10^{-18}$ | $1.0 \times 10^{-17}$<br>[$3.3 \times 10^{-16}$] | | Itikawa (2002)[&&&]<br>Orient and Srivastava (1987)[&&&&] |
| CO | | | [$2.5 \times 10^{-16}$] | | Orient and Srivastava (1987)% |
| OH | | $3.2 \times 10^{-18}$<br>$1.1 \times 10^{-18}$ | | | Chakrabarti et al. (2019)[%%]<br>Chakrabarti et al. (2019)[%%%] |
| NO | | | [$3.6 \times 10^{-16}$] | [$1.3 \times 10^{-16}$] | Iga et al. (1996) |
| $N_2O_5$ | | | | | Mason et al., (2015) |
| HCl | | | | | Hamada and Sueoka (1994) |
| $C_2H_6$ | | | $4.0 \times 10^{-19}$ | $5.4 \times 10^{-20}$ | Shirai et al. (2002)[%%%%] |
| $H_2S$ | | | [$8.8 \times 10^{-16}$] | | Zecca et al., (1992)[+] |
| $H_2CO$ | | | | | Vinodkumar et al., (2011) [++] |

Table A2a: Cross sections ($cm^2$) for neutral electronic dissociation ($\sigma_n$) (shown in black as used in our model BLACKWOLF) and for total electronic dissociation ($\sigma_t$) (shown in grey in square brackets) (i.e. including e.g. molecular ionization and ion dissociation; the grey values are shown for information

but they were not used in our (neutral) chemical network. Four electron energy (eV) ranges are shown, together with associated references for the gas phase species considered in this work. Energy intervals have centre point energies corresponding to 1, 10, 100 and 1000eV. Empty boxes in the Table indicate that data is not available. *Their Figure 3, filled black circles. **Their Figure 1 for $N_2^+$ formation by electron impact of $N_2$, filled black circles, data from 15 to 167 eV interpolated to the energy range indicated. ***Their Figure 3, continuous black line marked $\sigma_M$. ****Their Figure 3 for $O_2^+$ formation by electron impact of $O_2$, open circles, eV range as for $N_2$ above. #Ozone collision with low energy electrons (~1-2 eV) favors anionic dissociation to form $O_2^-$ and $O^-$ (see e.g. Matejcik et al., 1997, their Figure 4) rather than neutral dissociation products. ##Their Figure 2, upper panel. ###$N_2O$ favors formation of molecular and atomic cations upon collision with an electron, rather than neutral products. ####Their Table 1 for the total formation of $N_2O$ into $N_2O^+$, $N_2^+$, $NO^+$, $N^+$ and $O^+$. $CH_3Cl favors formation of $CH_3$ + $Cl^-$ rather than neutral dissociation. $$Their Figure 3 at 300K. Value is temperature-dependent and uncertain by several orders of magnitude. $$$Figure 1 of that work for the sum of the first and second ionization cross sections. $$$$Total cross section, their Table 2. §Formation of $CH_3$+H, their graph 16. §§Their Table 4 for total formation ($\sigma$(T)) of $CH_4^+$, $CH_3^+$, $CH_2^+$, $CH^+$ and $C^+$. §§§Their Figure 2, continuous red line. §§§§Their Figure 18 (open circles) and their Table 24 for formation of ground-state OH plus H. &Their Figure 18 (continuous, black line) for formation of ground-state $O(^1S)$ plus 2H. &&See Itikawa and Mason (2005) for example their Figure 12 for a range of cation products ($H_2O^+$, $OH^+$, $O^+$, $O^{++}$ and $H_2^+$). &&&Figure 19 from that work, producing $O(^1S)$. Note: we accordingly included into our model BLACKWOLF the collisional deactivation rate: $O(^1S)+CO_2 \rightarrow O(^1D)+CO_2$ with $k=2\times10^{-11}\exp(-1327/T)$ (Bhardwaj and Raghuram, 2012, their Table 2). %Their Figure 5, total cross section. %%Their Figure 5, green line forming $O(^3P)$ and $H(^2S)$. %%Their Figure 5, red line forming $O(^1D)$ and $H(^2S)$.). %%%%Their Figure 61. +Their Table 2. ++Their Figure 1.

Table A2a suggests that a range of species can favor ionization (shown in grey) over neutral dissociation (shown in black). Especially for polar molecules the more electronegative atomic constituent (denoted here as 'X') can favor electron attachment to the molecule followed by molecular dissociation to form the $X^-$ anion, a process termed dissociative electron attachment. Electron collision can also change the rotational and vibrational energetic states of target species and alter e.g. their photochemical responses. Cross sections for these processes, however, are generally not well known for numerous species and this is therefore a subject for future works.

**Caveats to Table A2a**

(1) Roldán et al. (2004) suggest cross sections for low energy electron-scattering cross sections could be over-estimated by ~40% due to simplifications involving the first Born approximation which assumes that the scattered wave function can be approximated by a plane wave.

(2) Regarding $N_2$ and $O_2$ air shower effects from dissociation of these species are typically formulated differently in Earth and Earth-like models. Starting with the IPP rate, the efficiency of NOx and HOx production per IPP is estimated e.g. from complex Earth chemical networks and inserted as production terms into the more straightforward chemical networks simulating Earth-like atmospheres (see e.g. Scheucher et al., 2018). This approach indirectly simulates the effect of chemical gas-phase pathways that produce NOx from atomic nitrogen (in the case of $N_2$) and a chemical cluster mechanism involving water molecules which overall produce HOx (in the case of $O_2$). In our work we do not include this IPP approach in BLACKWOLF since we wish to focus on the effect of neutral dissociation via electrons only.

(3) In our study we consider only the effect of neutral dissociation due to low energy electrons. However, other particles in the air shower can dissociate and ionize gas-phase species

although this effect is likely to be smaller in comparison. For example, Gobet et al. (2001) present ionization cross sections for $H_2O$ by low energy protons. Ben-Itzhak et al., (1994) present ionization and fragmentation of $CH_4$ by fast impact protons.

(4) In addition to electron-induced neutral and ionic dissociation gas phase species can also undergo collisional excitation e.g. to higher vibrational and electronic states which can impact photochemical reactions.

**Appendix 2b: Electron Fluxes**

The atmospheric flux profiles for the low energy (0.5-5000eV) secondary electrons were derived using the following two methods:

**Method 1** - used as input the medium energy electron flux model results from Ehresmann (2012) (who also used PLANETOCOSMICS) scaled linearly to the required low energies for the surface of Mars at solar minimum as shown in Figure A2c_1:

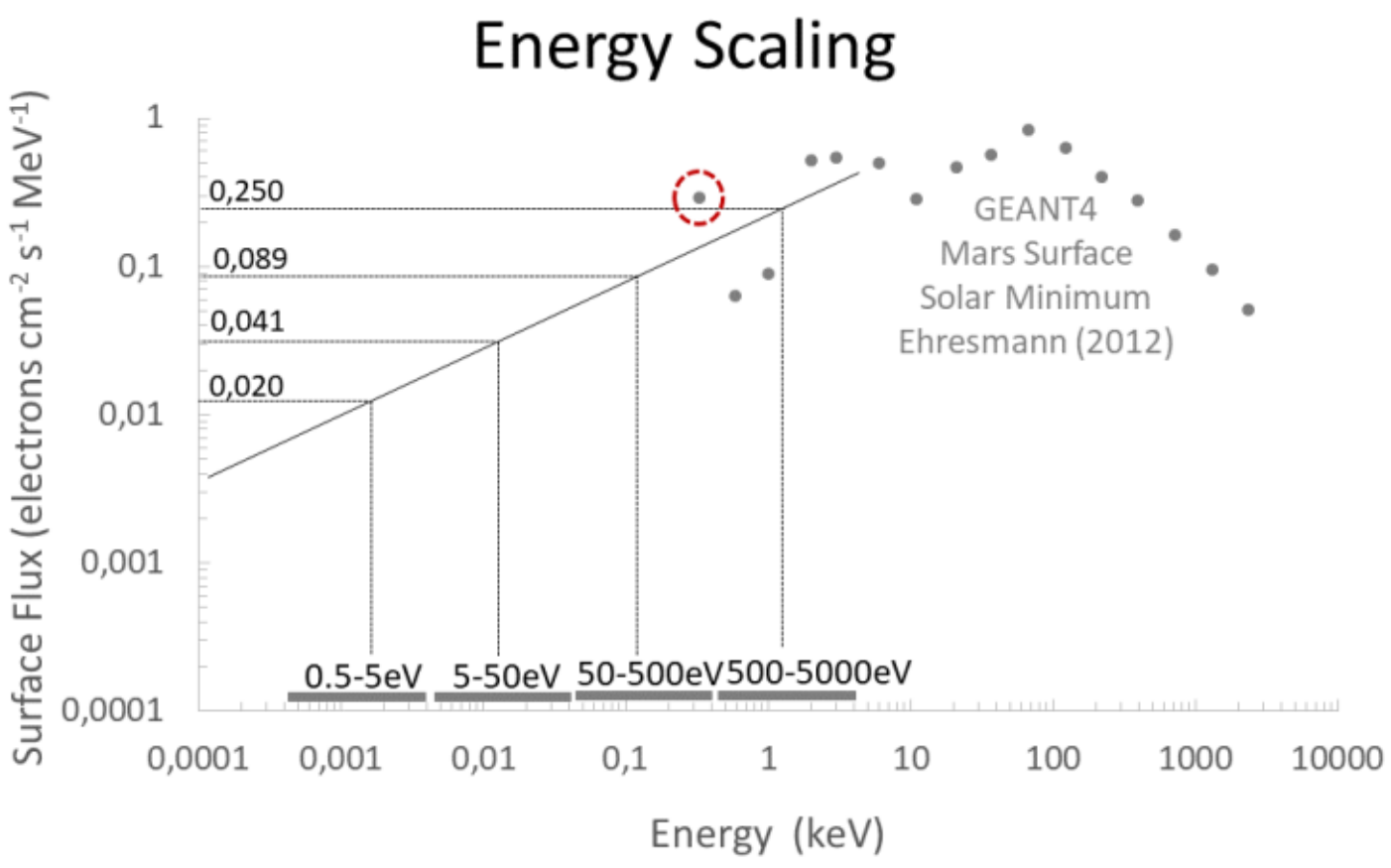


Figure A2c_1: Energy scaling to obtain low energy (0.5-5000) eV electron fluxes. Data shows output for the Martian surface at solar minimum (Ehresmann, 2012).

The pressure dependence for the low energy electrons is assumed to be similar to that of the IPP rates calculated in the PLANETOCOSMICS model runs for the Carrington-like event (see main text Figure 15).

Figure A2c_2 shows the pressure scaling:

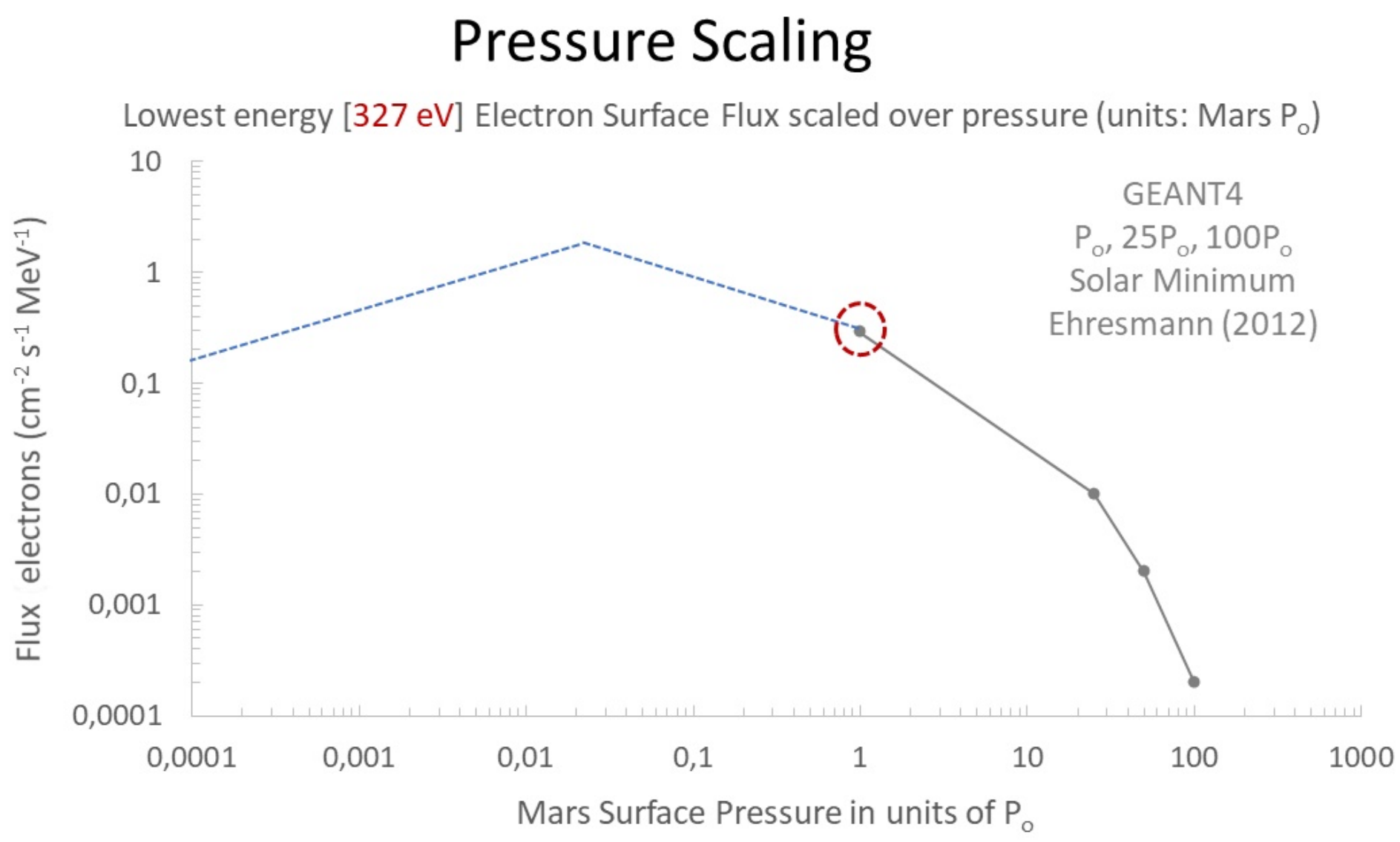


Figure A2c_2: Pressure scaling of low energy electron fluxes. Grey values show output from Ehresmann (2012). Dashed blue lines show pressure scaling derived such that the flux increases to a maximum at around 0.3hPa based on the IPP response (see Figure 15).

The final flux for the given electron energy interval and pressure is obtained by multiplying the electron flux (Figure A2c_2) at the given pressure level by the energy scaling factor, $F_e$ for the given energy interval (e.g. $F_e$=0.02 for the lowest interval in 0.5-5 eV as shown in Figure A2c_1) and by the pressure scaling factor shown in Figure A2c_2. The final flux profiles are shown in the main text as Figure 10.

**Method 2** – starting (as for method 1) with the electron fluxes in Figure A2c-2 we then multiply by the number of low energy electrons ($N_e$) produced by a single 10 GeV incoming proton calculated by the theoretical study of Paschalis et al. (2014) as a percentage of the total particles produced (their Figure 5, top right panel) at the top-of-atmosphere for modern Earth. Their results suggested $N_e$ (1eV) = 0.09%, $N_e$ (10eV) = 0.9%, $N_e$ (100eV) = 8% and $N_e$ (1000eV) = 30%. Finally, we multiply by the fraction (N_10 GeV / $N_{tot}$) where N_10 GeV is the number of incoming protons with energies in the 10 GeV bin from the Paschalis study and $N_{tot}$ is the total number of incoming protons at the Martian top-of-atmosphere for the GLE event calculated by PLANETOCOSMICS (see Herbst et al., 2019)[a]. For the pressure dependence through the atmosphere we assume the proton flux is reduced by a factor of 10 and 100 at 80 km and 0 km respectively.

**Appendix 2c – Implementation of electronic dissociation into BLACKWOLF**

We implemented the following energy dependent loss rate coefficients (with units $s^{-1}$ analogous to the photolysis coefficient) due to electronic neutral dissociation ($L_e$) summed over the four energy (E) intervals taken from Table A2a:

$$L_e = \Sigma_{E=1,4}\, \sigma(E) \cdot F_e(z) \cdot h$$

where $\sigma(E)$ denotes the energy-dependent cross section ($cm^2$) for neutral dissociation by low energy electrons for the energy intervals shown in Figure A2c_1, $F_e(z)$ is the altitude (z) dependent flux ($cm^{-2}$ $s^{-1}$) of the secondary electrons and h= model (BLACKWOLF) vertical grid height (km).

**References**

Aoki, S., et al., A&A, 610, A78 (2018)
Aoki, S., et al., Geophys. Res. Lett., 48, e2021GL092506 (2021)
Banjac, S., et al., J. Geophys. Res., https://doi.org/10.1029/2018JA026042 (2019)
Ben-Itzhak, I., et al., Phys. Rev. A 49, 881 (1994)
Bhardwaj, A., and Raghuram, S. ApJ, 748, 1-18 (2012)
Blamey, N. J. F., et al., Nature Comm., 6, 7399 (2015)
Bouche, J., et al., 238, JQSRT, 106498 (2019)
Bougher, S. W., et al., J. Geophys. Res. Planets, 120, 2, doi:10.1002/2014JE004715 (2015)
Boxe, C. S., et al., Icarus, 242, 97-104 (2014)
Brasseur, G.P., Orlando, J. J., and Tyndall, G. S., (eds.), Atmospheric Chemistry and Global Change. Oxford University Press, New York (1999)
Brehm, N., et al., Nat. Geosci., 14, 10-15 (2021)
Buford Price P., PSS, 58, 1199-1206 (2010)
Burkholder, J. B., et al., Chemical kinetics and photochemical data for use in atmospheric studies; evaluation number 19. http://hdl.handle.net/2014/49199 (2020)
Burnett, E. B., and Burnett, C. R., J. Atmos. Chem. 21, 13-41 (1995)
Cardnell, S., et al., J. Geophys. Res. Planets, 121, 11, doi:10.1002/2016JE005077 (2016)
Chakrabarti et al., Plasma Sci. Tech.,28, 085013, 1-8 (2019)
Chassefière, E., Icarus, 204, 137-144 (2009)
Chastain, B. K., and Chevrier, V., PSS, 55, 1246-1256 (2007)
Civiš, S., et al., Earth Spa. Chem., 3, 221-222 (2019)
Cosby, P., J. Chem., Phys., 98, 9544-9553 (1993)
Court, R. W., and Sephton, M. A., Earth Plan. Sci. Lett., 288, 382-385 (2009)
Daerden, F., et al., Icarus, 326, 197-224 (2019)
Desorgher, L. 2005, PLANETOCOSMICS Software User Manual, Tech. Rep. 1, University Bern
Dunlop, J. R., and Tully, F. P., J. Phys. Chem., 97, 11,148-11,150 (1993)
Ehresmann, B., The Martian Radiation Environment - Early Mars and Future Measurements with the Radiation Assessment Detector, PhD Dissertation, Christian-Albrechts-Universität zu Kiel (2012)
Escamilla-Roa, E., et al., Icarus, 153, 163-171 (2018)
Etiope, G., et al., Science Rep., 8, 8728 (2018)
Fedorova, A., J., et al., Geophys. Res. Planets, doi: 10.1029/2020JE006616 (2020, accepted)
Farrell, W. M., et al., Geophys. Res. Lett., 33, 21 (2006)
Fichtner, H., et al., Solar Activity, the Heliosphere, Cosmic Rays and Their Impact on the Earth's Atmosphere System, 55-78, In: Climate and Weather of the Sun-Earth System (CAWSES): Highlights from a Priority Program"", Ed., F. J. Lübken, Springer Netherlands, Dordrecht (2012)
Fonti, S., and Marzo, G. A., A&A, 512, A51 (2010)
Fonti, S., et al., A&A, 581, A136 (2015)
Formisano, V., et al., Science, 306, 1759-1761 (2004)
Fries et al., Geochem. Persp. Lett., 3, (2016)
Geminale, A., et al., PSS, 59, 137-148 (2011)
Gierasch, P., and Conrath, B., Recent Adv. in Planet. Met., ed. G. E. Hunt (Cambridge Univ. Press), 121 (1985)
Gillen, E., et al., Icarus, 336, 113407 (2020)
Giuranna, M., et al., Nature Geoscience, 12, 326-332 (2019)
Gobet, F., et al., Phys. Rev. Lett., 86, 17, 3751-3754 (2001)

González-Galindo, F., et al., J. Geophys. Res. Planets, 118, 10, doi:10.1002/jgre.20150 (2013)
Gronoff, G., et al., Adv. Spa. Res., 55, 1799-1805 (2015)
Guo, J., et al., J. Spa. Weather Spa. Clim., 9, A2, (2020)
Haberle, R., et al., The Atmosphere and Climate of Mars, Cambridge University Press (2017)
Hamada, A., and Sueoka, O., Int. Phys. B. 27, 20 (1994)
Hartogh, P., et al., A&A, 521, A49 (2010)
Herbst, K., et al., J. Geophys. Res., 122, 23-34 (2017)
Herbst, K., Interaction of Cosmic Rays with the Earth's Magnetosphere and Atmosphere - Modeling the Cosmic Ray Induced Ionization and the Production of Cosmogenic Radionuclides, PhD Thesis, Christian-Albrechts-Universität zu Kiel (2013)
Herbst, K., et al., Ann. Geophys., 115, 1637-1643 (2013)
Herbst, K., et al., J. Geophys. Res., 115, D1 (2017)
Herbst, K., et al., A&A, 631, A101 (2019a)
Herbst, K., et al., A&A, 624, A124 (2019b)
Herbst, K., et al., A&A, 633, A15 (2020)
Holmes, J. A., et al., Geophys. Res., Lett., 44, 16, 8611-8620 (2017)
Iga, I., et al., J. Geophys. Res. Planets, 101, E4, doi: 10.1029/96JE00467 (1996)
Itikawa, I., J. Phys. Chem. Ref. Data, 31, 749 (2002)
Itikawa, Y., and Mason, N., J. Phys. Chem. Ref. Data, 34, 1 (2005)
Jensen, S. J. K., et al., Icarus, 236, 24-27 (2014)
Keppler, F., et al., Nature, 486, 93-96 (2012)
Knutsen, E. W., et al., Icarus, 357, 114266 (2021)
Köhler, J., et al., Ann. Geophys., 34, 133-141 (2016)
Komatsu, G., et al., PSS, 59, 169-181 (2011)
Korablev, O., et al., Nature, 568, 517-520 (2019)
Korablev, O., et al., Science Adv., 7, 7, 1-8 (2021)
Krasnopolsky, V. A., et al., Geophys. Res. Abs., 6, 06169 (2004)
Krasnopolsky, V. A., et al., Icarus, 182, 80-91 (2006)
Krasnopolsky, V. A., et al., Icarus, 190, 93-102 (2007)
Krasnopolsky, V. A., et al., Icarus, 207, 638-647 (2010)
Krasnopolsky, V. A., and Lefèvre, F., Comparative climatology of terrestrial planets, "Chemistry of the atmospheres of Mars, Venus and Titan" p. 231-275, University of Arizona Press (2013)
Krasnopolsky, V. A., Spectroscopy and photochemistry of planetary atmospheres and ionospheres, chapter 10: Photochemical Modeling, Cambridge Planetary Science (2019)
Kress, M. E., and McKay, C. P., Icarus, 168, 475-483 (2004)
Kumar, R., Chem. Phys. Lett., 609, 108-112 (2014)
Lefèvre, F., et al., J. Geophys. Res. Planets, 109, E7, doi: 10.1029/2004JE002268 (2004)
Lefèvre, F., and Forget, F., Nature, 460, 720-723 (2009)
Lefèvre, F., and Krasnopolsky, V., Chapter 13 "Atmospheric Photochemistry" in "The atmosphere and climate of Mars", Eds., Haberle, R. M., Clancy, R. T., Forget, F., Smith, M. D., and Zurek, M W., Cambridge University Press (2017)
Lefèvre, F., Chapter 12 "The enigma of methane on Mars", in "Biosignatures for Astrobiology", Eds. Cavalazzi, B., and Westall, F., Springer Nature Switzerland AG (2019)
Lehmann, R., J. Atmos. Chem., 47, 45-78 (2004)
Lindner, B. L., PSS, 26, 125-144 (1988)
Liuzzi, G., et al., Icarus, 321, 671-690, (2019)

Lo, D. Y., et al., Icarus, 352, 114001 (2020)
Lyons, J. R., et al., Geophys. Res. Lett., 32, L13201 (2005)
Maeda, K., and Aikin, A. C., PSS, 371-384 (1968)
Märk, T. D., J. Chem. Phys., 63, 3731-3736 (1975)
Mason, N. J., et al., RAS 357, 1125 (2015)
Matejcik, S., et al., Plasma Source Sci. Tech., 6, 140 (1997)
Matthiä, D., et al., J. Geophys. Res., 114, A08104, (2009)
Matthiä, D., and Berger, T., Life Sci. Spa. Res., 55, 14, 57-63 (2017)
Michaels, T. I., and Rafkin, S, C. R., QJRMS, 130, 599 (2004)
Molina-Cuberos, G. J., et al., Icarus, 27, 1801-1806 (2001)
Molina-Cuberos, G. J., et al., J. Geophys. Res. Planets, 107, E5 (2002)
Montmessin, F., et al., A&A. 650, A140 (2021)
Moores, J. E., et al., PSS, 147, 48-60 (2017)
Moores, J. E., et al., Nature Geoscience., 12, 321-325 (2019)
Mouden, Y., and McConnell, J. C., Icarus, 188, 18-34 (2007)
Mousis, C., et al., Icarus, 278, 1-6 (2016)
Mumma, M. J., et al., Science, 323, 1041-1045 (2009)
Nagy, T., et al., Combustion and Flame, 162, 2059-2076 (2015)
Nair, H., Icarus, 111, 124-150 (1994)
Nixon, M. L., et al., News Rev. Astron. Astrophys., 54, 13-16 (2013)
Oehler, D. Z., and Etiope, G., Astrobiol., 17, 12, doi: 10.1089/ast.2017.1657 (2017)
Olsen, K. S., et al., A&A 639, A141 (2020)
Olsen, K. S., et al., A&A, doi:10.1051/0004-6361/202140329 (2021)
Orient, O. J., and Srivastava, S. K., J., Phys. B: Atomic Mol. Opt. Phys., 20, 3923-3936 (1987)
Owen, T., et al., J. Geophys. Res.,82, 4635-4639 (1977)
Oze, C., and Sharma, M., Geophys. Res., Lett., 32, L10293 (2005)
Padovani, M., et al., A&A, 619, A144, 1-7 (2018)
Paschalis, P., et al., New Astron., 33, 26-37 (2014)
Pearl, D. M., J. Chem. Phys., 102, 2737-2743 (1995)
Pla-Garcia, G., et al., J. Geophys. Res. (Planets), 124, 2141-2161, (2019)
Rangwala, S. A., et al., J. Phys. B: Atomic Mol. Opt. Phys., 32, 3795-3804 (1999)
Raukunen, O., et al., J. Spa. Weather Spa. Clim., 8, A04, (2018)
Risse, M., and Heck, D., Astroparticle Phys., 20, 661-667 (2004)
Roldán, A., et al., J. App. Phys., 95, 5865-5870 (2004)
Roos-Serote et al., J. Geophys. Res., Planets, 121, 2108-2119 (2016)
Röstel, L., et al., J. Geophys. Res., (Planets), 125, 1-12, e2019JE006246 (2020)
Saunois, S. M., Bousquet, P., Poulter, B., et al., Earth Sys. Sci. Data, 8, 697-751 (2016)
Scheucher, M., et al., ApJ, 863, 1 (2018)
Scheucher, M., et al., ApJ, 898, 44, 1-19 (2020)
Schuerger, A. C., et al., J. Geophys. Res., 117, E08007 (2012)
Shirai, T., et al., Atomic Dat. Nuc. Dat. Tab., 80, 147-204 (2002)
Sholes, S. F., et al., Astrobiol., 19, 655-668 (2019)
Smith, M. D., et al., Icarus, 301, 117-131 (2018)
Stock, J. W., et al., Icarus, 219, 13-24 (2012)
Stock, J. W., et al., Icarus, 291, 192-202 (2017)
Summers, M. E., et al., Geophys. Res. Lett., 20 (2002)
Thomas, C., et al., PSS, 57, 42-47 (2009)
Trokhimovskiy, A., et al., Icarus, 251, 50-64 (2015)

Usoskin, I. G., and Kovaltsov, G. A., Geophys. Res., Lett., https://doi.org/10.1029/2021GL094848 (2021)
Viscardy, S., et al., Geophys. Res. Lett., 43, 1868-1875 (2016)
Villanueva, G. L., et al., Icarus, 233, 11-27 (2013)
Villanueva, G., L., et al., Science Adv., 7, eabc8843 (2021)
Vinodkumar, M., et al., Int. J. Mass Spec., 308, 35-40 (2011)
Viúdez-Moreiras, D., et al., Geophys. Res. Lett., 47, 3, doi:10.1029/2019GL085694 (2020)
Webster, C., et al., Science, 347, 415-417 (2015)
Webster, C., et al., Science, 360, 1093-1096 (2018)
Webster, C., et al., A&A, 641, L3 (2020)
Webster, C. R. et al., Astronomy & Astrophysics 650, A166 (2021)
Westall, F., et al., Astrobiology, 15, 998-1029 (2015)
Williams, D., Mars Fact Sheet (Greenbelt, MD: National Space Science Data Center), https://nssdc.gsfc.nasa.gov/planetary/factsheet/marsfact.html (2010)
Wunderlich, F., et al., ApJ, 901, 126, 1-31 (2020)
Wong, A. S., et al., Adv. Spa. Res. 33, 2236-2239 (2004)
Yung, Y., et al. Astrobiology, 18, 1221-1242 (2018)
Zahnle, K., et al., Icarus, 212, 493-503 (2011)
Zecca, A., et al., Phys. Rev. A., 45, 2777-2783 (1992)